\documentclass{ametsocv6.1}
\usepackage{graphicx}
\usepackage{placeins}
\usepackage{todonotes}
\usepackage{textgreek}
\usepackage{mathrsfs} 
\usepackage{bm}
\usepackage{enumitem}

\usepackage[normalem]{ulem}
\usepackage{amsmath}
\usepackage{amssymb}
\usepackage{mathtools}
\usepackage{mathptmx}
\usepackage{lineno}
\nolinenumbers
\usepackage{fix-cm} 
\DeclareMathAlphabet{\mathsfit}{T1}{\sfdefault}{\mddefault}{\sldefault}
\SetMathAlphabet{\mathsfit}{bold}{T1}{\sfdefault}{\bfdefault}{\sldefault}

\newcommand\Ro{\mbox{\textit{Ro}}} 

\def\ee{{\rm e}}
\def\ii{{\rm i}}

\newcommand{\changefunction}[1]{%
  \expandafter\renewcommand\csname#1\endcsname[1][]%
    {\qopname\relax o{#1}\ifx\relax##1\relax\else^{##1}\fi}}
\changefunction{sin}
\changefunction{cos}
\changefunction{tan}

\renewcommand{\arctan}{\tan[-1]}

\usepackage{bigstrut}
\usepackage{booktabs, multirow} 
\usepackage{soul}
\usepackage{changepage,threeparttable} 
\usepackage{tabularx,ragged2e,booktabs}

\title{Spontaneous emission of internal waves by a radiative instability}

\authors{
Subhajit Kar
\correspondingauthor{
Subhajit Kar, subhajitkar@mail.tau.ac.il
}
\aff{a}, 
Roy Barkan\aff{a,b},
James C. McWilliams\aff{b} 
and M. Jeroen Molemaker\aff{b}
}

\affiliation{
\aff{a}Porter School of the Environment and Earth Sciences, Tel Aviv University, Ramat Aviv,  Israel 6997801.
\\
\aff{b}Department of Atmospheric and Oceanic Sciences, University of California, Los Angeles, CA, USA.
}

\abstract{
The spontaneous emission of internal waves (IWs) from balanced mesoscale eddies has been previously proposed to provide a source of oceanic IW kinetic energy (KE). 
This study examines the mechanisms leading to the spontaneous emission of spiral-shaped IWs from an anticyclonic eddy with an order-one Rossby number, using a high-resolution numerical simulation of a flat-bottomed, wind-forced, reentrant channel flow configured to resemble the Antarctic Circumpolar Current.
It is demonstrated that IWs are spontaneously generated as a result of a
loss of balance process that is concentrated at the eddy edge, and then radiate radially outward. A 2D linear stability analysis of the eddy shows that the spontaneous emission arises from a radiative instability which involves an interaction between a vortex Rossby wave supported by the radial gradient of potential vorticity and an outgoing IWs. This particular instability occurs when the perturbation frequency is superinertial. This finding is supported by a KE analysis of the unstable modes and the numerical solution, where it is shown that the horizontal shear production provides the source of perturbation KE. Furthermore, the horizontal length scale and frequency of the most unstable mode from the stability analysis agree well with those of the spontaneously emitted IWs in the numerical solution. 
}

\begin{document}
\maketitle

%
%
%
\statement
Spontaneous emission of internal waves (IWs) describes a process by which a oceanic large-scale and slow currents can spontaneously emit IWs. Recent observations and numerical studies suggest that spontaneous IW emission can provide an important IW energy source.  Identifying the mechanisms responsible for spontaneous IW emission are thus of utmost importance, because IW breaking has crucial effects on the oceanic large scale circulation. 
 In this study, we examine the spontaneous emission of IWs from a numerically simulated anticyclonic eddy. We show that the emission process results from a radiative instability that occurs when the frequency of the perturbation is larger than the Coriolis frequency. This instability mechanism can be significant across the oceans for flow structures with order-one Rossby numbers (a measure of the flow nonlinearity). 



%
%

%

\section{Introduction}
Internal waves (IWs) are ubiquitous in the ocean
 and their breaking drives turbulent mixing that shapes large-scale circulation patterns and the distribution of heat and carbon in the climate system 
 \citep{munk1998abyssal,whalen2020internal}.
 They represent a large energy reservoir, with about 1TW converted from barotropic tides \citep{egbert2000significant,nycander2005generation}, and another 0.3-1.4 TW converted into near-inertial IWs, mainly from high-frequency wind forcing \citep{alford2003improved,rimac2013influence}. 
 
 Another possible IW generation mechanism that has been proposed is termed spontaneous emission \-- a process describing the spontaneous generation of IWs from so called balanced motions \citep[see][and refrences therein]{vanneste2013balance}. 
 These balanced motions satisfy the \textit{invertibility principle of Potential vorticity} (PV) \--- at a given  instant, all dynamical fields (e.g., velocity, density) can be deduced by inverting the PV without the need to  time evolve each of the fields separately \citep{hoskins1985use}. A classical example is the quasigeostrophic (QG) model \citep{pedlosky2013geophysical} that is quite successful in describing the dynamics of oceanic mesoscale eddies; typically characterized by small Rossby numbers ($Ro \ll 1$) and large Richardson numbers ($Ri \gg 1$).



\cite{ford1994gravity} and \cite{ford2000balance} demonstrated the analogy between spontaneous emission of IWs from a balanced flow and Lighthill radiation of acoustic wave from a turbulent flow \citep{lighthill1954sound}. \cite{vanneste2004exponentially} and \cite{vanneste2008exponential} showed that in the low-$Ro$ regime spontaneous emission is expected to be exponentially small. 
Conversely, \cite{williams2008inertia} found in laboratory experiments that the amplitude of the spontaneously emitted IWs depends linearly on $Ro$. Under both paradigms, these previous findings suggest that spontaneous emission could be significant in high-$Ro$ flows.

Indeed, \cite{shakespeare2014spontaneous}  showed analytically that the spontaneous emission from strained fronts can be significant for large strain values, representative of an $O(1)$ Rossby number regime. 
Later, \cite{nagai2015spontaneous} performed an idealized simulation of a Kuroshio front and demonstrated significant spontaneous emission of IW energy from the front. 
The emitted IWs were eventually reabsorbed into the mean flow at depth, thereby providing a redistribution of balanced flow energy rather than a pure sink. 
Using high-resolution numerical simulations of an idealized channel flow, \cite{shakespeare2017spontaneous} also reported spontaneous emission of IWs from surface fronts, which were further amplified at depth through energy exchanges with the mean flow. 



Direct observational evidence of spontaneous emission in the ocean is scarce, likely because of the difficulty in eliminating other IW generation mechanisms using sparse measurements. 
\cite{alford2013observations} measured the rate of generation of IWs from a subtropical frontal jet in the Northern Pacific Ocean to be $0.6-2.4$ mW m$^{-2}$, which leads to a source of about $0.2-0.9$ TW IW energy, when extrapolated to the global ocean. This rough evaluation 
is comparable to the estimate of wind-forced near-inertial IWs, thereby suggesting that spontaneous emission could be significant to the ocean's KE budget. 
\cite{johannessen2019observations} also
showed evidence of spontaneous emission of IWs from a mesoscale, baroclinic anticyclonic eddy in the Greenland sea (at latitude of  $\sim 78^{\circ}$N) with horizontal scale of $1$km.
Moreover, using Synthetic Aperture Radar measurements,  \cite{chunchuzov2021possible} observed the emission of spiral-shaped IWs of horizontal scale of $0.4-1$km from the edge of a high-$Ro$ submesoscale cyclonic eddy near the Catalina Island. 

In this article, we investigate the spontaneous emission of spiral-shaped IWs from an anticyclonic eddy of $\mathcal{O}(1)$ Rossby number, using a high-resoluion numerical simulation of a statistically equilibrated channel flow. 
We show that the spontaneous emission is directly  linked to a loss of balance (LOB) process that results from a radiative instability of the eddy. To our knowledge this is the first demonstration of such instability mechanism in forced dissipative numerical solutions. 

The article is organized as follows: In section \ref{setup}, we describe the numerical setup used to study the spontaneous emission of IWs from the eddy. The quantification of LOB of the mean flow and the generation and propagation of  the radiated IWs are discussed in section \ref{lob}. In section \ref{mechanisms}, we examine possible mechanisms leading to the LOB and spontaneous emission. The setup and methodology used to carry out a 2D linear stability analysis of the eddy circulation is described in section \ref{stability_analyis}. In section \ref{instab}, we present the results of the stability analysis and compare them with the numerical solution. The instability mechanism is discussed in section \ref{discussion}, and in 
section \ref{summary} we summarize our findings and their implications for realistic ocean scenarios.

\section{Numerical setup}
\label{setup}
The numerical simulations are performed using flow\_solve \citep{winters2012modelling}, a pseudospectral, non-hydrostatic, Boussinesq solver. The setup consists of a reentrant channel flow on an $f$-plane over which wind blows to mimic an idealized configuration of the Antarctic Circumpolar Current (ACC), with an initial stratification profile based on observations from the Southern  Ocean \citep{garabato2004widespread}. Without loss of generality, the Coriolis frequency $f >0$ and the value is fixed to $f=1.2\times10^{-4}$ s$^{-1}$. The domain size in the zonal, meridional, and vertical $(\hat{x}, \hat{y}, \hat{z})$ are $L_x=200$ km, $L_y=200$ km, and $H=2$ km, respectively. The boundary conditions are periodic in the zonal direction, free-slip wall in the meridional direction, and free-slip rigid lid in the vertical direction.

The numerical analysis shown in this manuscript is based on one of the simulations previously discussed in \cite{barkan2017stimulated}. The simulation is forced by a steady wind stress $\tau_s$
of the form 
\begin{align}
\label{wind_stress}
    \tau_s(y) = \tau_0 \sin^2{\Big(\frac{\pi y}{L_y}\Big)} \hat{x},
\end{align}
where $\rho_0 \tau_0=0.1$ Nm$^{-2}$ and the reference density $\rho_0=10^3$ kgm$^{-3}$. 
The wind stress is applied as a body force confined to the upper $\sim 80$m, representing an effective mixed layer depth. This wind forcing drives a zonal jet (i.e., an idealized `ACC') and induces Ekman upwelling and downwelling that tilt the initially flat isopycnals, leading to baroclinic instability and the subsequent formation of mesoscale baroclinic vortices.

\begin{figure}
    \centering
    \includegraphics[width=\textwidth]{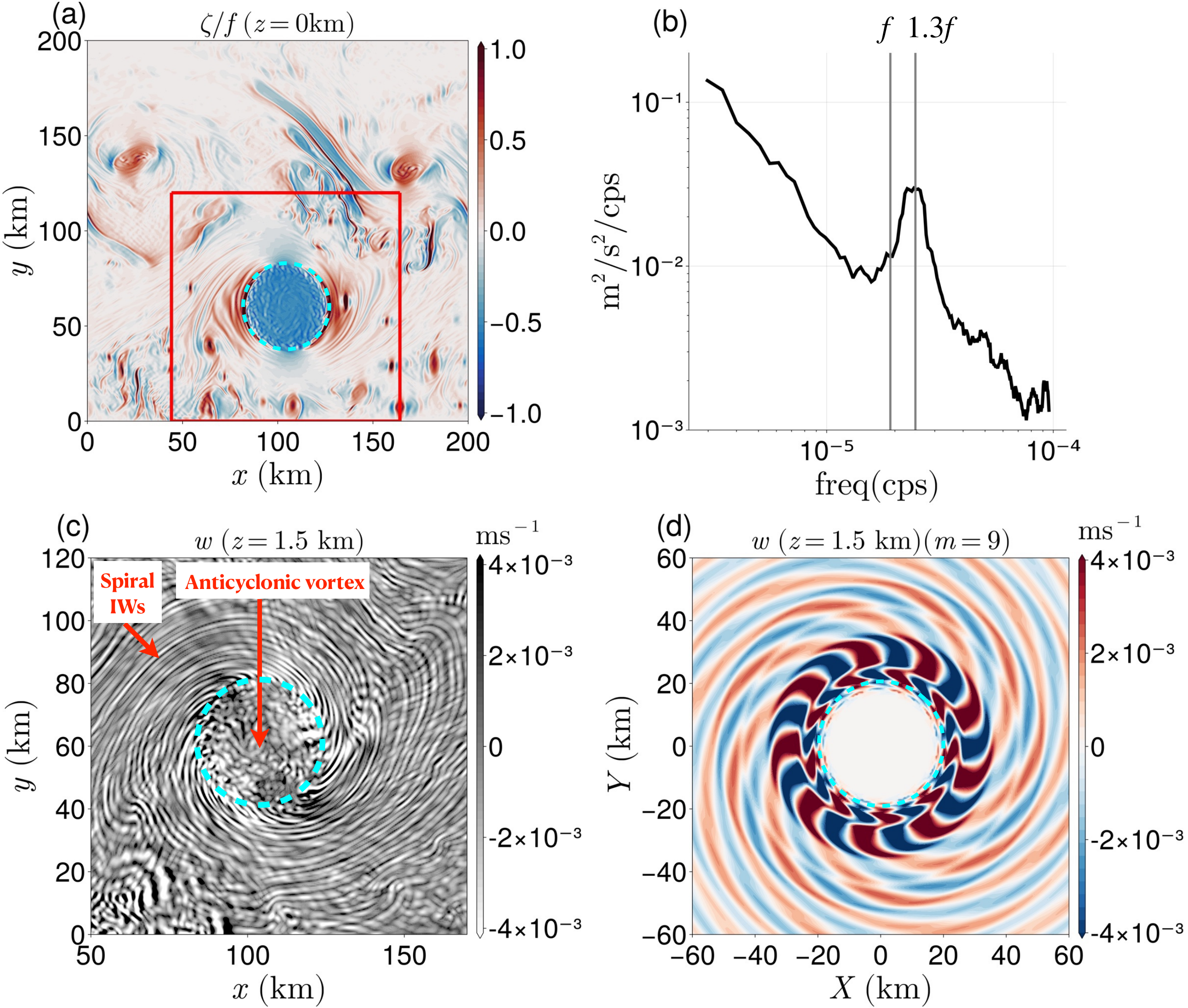}
    \caption{(a) A representative surface snapshot of  the vertical component of vorticity $\zeta$ (normalized by $f$). The red box region of size $(120 \text{km} \times 120 \text{km})$ is used to analyze IW generation and propagation from the anticyclonic eddy. (b) Vertical velocity $w$ frequency spectra in a frame moving with the ACC (Eq. \ref{ref_frame}). The spectrum is computed in the red box region shown in panel (a), excluding the anticyclonic eddy region. The spectrum peaks approximately at $1.3f$. (c) A representative snapshot of the vertical velocity $w$ at $z=1.5$km at the same time instance of panel (a). The spiral-shaped IWs radiated from the edge of the eddy are visible. Reflection of the radiated IWs from the free-slip wall at $y=0$km is also visible. The dashed cyan lines in panels (a), (c), and (d) mark the radius $R = 20$ km of the  eddy. Typically horizontal length scale of the emitted IWs is $\sim 4$ km.
    (d) Solution of vertical velocity $w$ obtained from $2$D linear stability analysis of the eddy for the case of azimuthal wavenumber $m=9$ (see Section \ref{stability_analyis} for more details). }
    \label{fig:fig1}
\end{figure}

A representative snapshot of the vertical component of vorticity at the surface shows a large anticyclonic eddy, two  cyclonic eddies, and smaller scale fronts and filaments with $O(1)$ Rossby numbers (Fig. \ref{fig:fig1}(a)). The corresponding vertical velocity in the vicinity of the anticyclonic eddy at 500m depth shows spiral-shaped structures that originate near the edge of the eddy (Fig. \ref{fig:fig1}(c)). 
The associated power spectral density of the   vertical velocity suggests that the spiraling structures may be the signature of spontaneously emitted IWs with an $\approx 1.3f$ frequency (Fig. \ref{fig:fig1}(b)).
In what follows, we will investigate in detail the mechanisms leading to the spontaneous IW emission from this anticyclonic eddy.


Because the anticyclonic eddy is being translated by the idealized `ACC'  in the $x$ direction with a nearly constant speed of $U_\text{ref} = 0.26$ ms$^{-1}$, we carry out the analysis that follows in the `ACC' reference frame 
\begin{align}
\label{ref_frame}
    X = x - U_\text{ref} t , \,\,\,\,\ Y=y, \,\,\,\,\,\ Z =z,
\end{align}
where $(x,y,z)$ are the Cartesian coordinates of the numerical simulation.


Furthermore, to separate the spontaneously emitted IWs from the slowly evloving mean flow, we decompose any field $\phi$ viz.
\begin{align}
\label{filter}
    \phi = \overline{\phi} + \phi^\prime, 
\end{align}
where the overline denotes a low-pass sixth-order Butterworth temporal filter with a frequency cutoff of $0.8f$, and the prime denotes the reminaing IW field. 
The filtering is applied in the moving reference frame $(X,Y,Z)$ to reduce the Doppler shifting effects \citep[e.g.,][]{rama2022importance}.

Throughout the article, we used the notation $\langle \, \rangle$ to represent an average quantity, and
subscript of the notation denotes the average along that direction unless otherwise stated, for example
\begin{align}
    \langle \phi \rangle_z = \frac{1}{H} \int_0^H \phi \, dz
\end{align}
denotes vertical average of $\phi$.



\section{Evidence of loss of balance and spontaneous emission}
\label{lob}
\begin{figure}
    \centering
    \includegraphics[width=\textwidth]{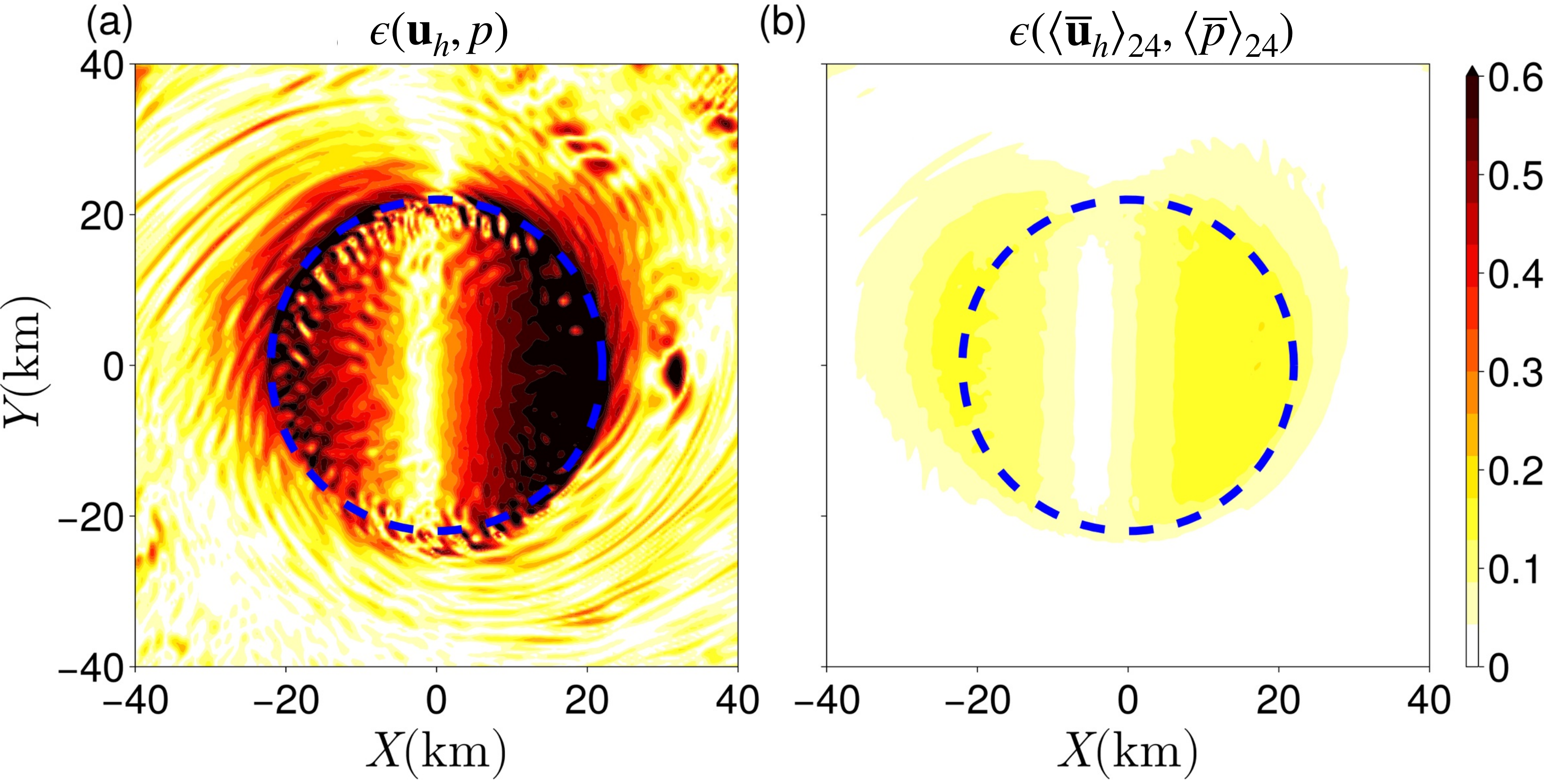}
    \caption{(a) A representative surface snapshot of the loss of balance parameter $\epsilon$ and (b) based on a daily time average of low-pass velocity and pressure field $(\langle \overline{\bm{u}}_h \rangle_{24}, \langle \overline{p} \rangle_{24})$  given by Eq. (\ref{degree_unbal}). 
    The daily time average is used to smooth out any small-scale motions within the eddy that cannot be removed by the Eulerian temporal filter.
    The dashed blue line marks the edge of the anticyclonic eddy. 
    }
    \label{fig:fig3}
\end{figure}

\begin{figure}
    \centering
    \includegraphics[width=\textwidth]{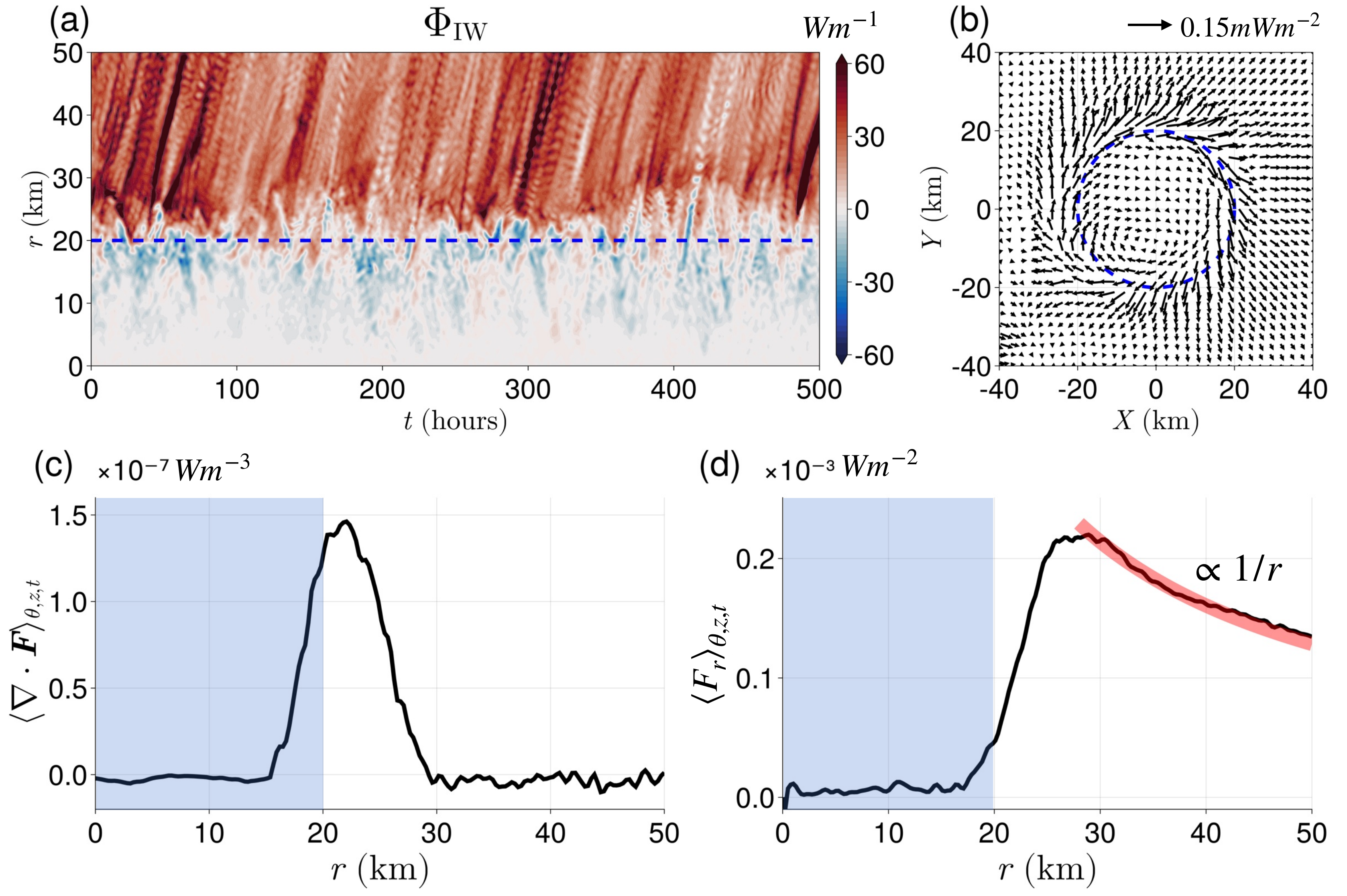}
    \caption{(a) Radial and time series plot of the IW energy propogation $\Phi_\text{IW}$  (Eq. \ref{energy_flux}). 
    (b) Time and vertically averaged IW energy flux vector $\bm{F}$ given by Eq. (\ref{flux_def}).
    The dotted blue lines
    in panels(a,b) indicate the edge of the anticyclonic eddy at  $r=20$ km. Azimuthal, vertical, and time-averaged of (c) divergence of the energy flux $\bm{F}$ given by Eq. (\ref{div_flux}) and (d) radial energy flux $F_r$. The thick red line shows the curve $r^{-1}$. The blue shaded regions in panels (c) and (d) show the eddy region. 
    The time averaging for 
    panels (b-d) is performed over $35$ inertial periods. 
    }
    \label{fig:fig2}
\end{figure}

To determine whether the IW signatures shown in Fig. \ref{fig:fig1}(c) are indeed associated with a loss of balance (LOB) in the anticyclonic eddy, 
we diagnose 
the departure from the gradient wind balance
\citep{mcwilliams1985uniformly} 
\begin{align}
\label{balance_motion}
-\nabla_h \cdot (\bm{{u}}_h \cdot \nabla_h \bm{{u}}_h) + f {\zeta} = \nabla_h^2 {p}, 
\end{align}
where $\nabla_h =(\partial_X, \partial_Y)$ is the horizontal gradient operator, ${\bm{u}}_h= (u,v)$ is the horizontal velocity vector, and ${p}$ is the pressure. 
The associated LOB measure for a given flow field $({\bm{u}}_h, {p})$ can be defined as  
\citep{capet2008mesoscale},
\begin{align}
\label{degree_unbal}
    \epsilon({\bm{u}}_h, {p}) = \frac{|\nabla_h \cdot (\bm{{u}}_h \cdot \nabla_h \bm{{u}}_h) - f {\zeta} + \nabla_h^2 {p}|}{|\nabla_h \cdot (\bm{{u}}_h \cdot \nabla_h \bm{{u}}_h)|+f|{\zeta}|+|\nabla_h^2 {p}| + \mu},
\end{align}
where the term $\mu= f{\zeta}_\text{rms} + (\nabla_h^2 {p})_\text{rms}$ is added to the denominator of Eq. (\ref{degree_unbal}) to 
eliminate the possibility of identifying weak flow regions as significantly unbalanced. 
The value of $\epsilon$ varies from $0$ to $1$, with $\epsilon=0$ ($\epsilon=1$) denoting fully balanced (unbalanced) motions.
A representative snapshot of $\epsilon$ at the surface shows significant imbalance around the edge of the anticyclonic eddy (Fig. \ref{fig:fig3}(a)). 
Evidently, the motions leading to loss of balance are quite rapid because the daily averaged low-pass velocity field is largely balanced (Fig. \ref{fig:fig2}(b)).
Hereinafter we refer to this balanced flow as the  \textit{mean flow} or \textit{basic state}. To denote it, we used subscript `$m$', which describes the daily average of the low-pass field.

To establish the connection between the rapid motions leading to LOB at the edge of the anticyclonic eddy and the spontaneous emission of IWs we first compute the  
 IW energy flux 
 \begin{align}
\label{flux_def}
    \bm{F} = \overline{\bm{u}^\prime p^\prime},
\end{align}
where $\bm{u}^\prime \equiv (u_r^\prime, u_\theta^\prime, w^\prime)$ and $p^\prime$ denote the IW velocity and pressure fields, respectively.
These IW fluxes are computed in  a cylindrical coordinate system $(r, \theta, z)$ centered around the anticyclonic eddy, with
\begin{align}
\label{cylindrical_coordinates}
    r =  \sqrt{X^2 +Y^2}, \,\,\,\, \theta = \arctan(Y/r).
\end{align}
The temporal filter (Eq. \ref{filter}) is applied after removing the depth averaged fields at each time instant. 

The associated outward propagating IW energy 
can be estimated using the azimuthally- and vertically averaged radial energy flux viz. \citep{voelker2019generation} 
\begin{align}
\label{energy_flux}
    \Phi_\text{IW}(r, t) = \frac{1}{H} \int_0^H \int_0^{2\pi} F_r r  d\theta dz,
\end{align}
where $F_r=\overline{u_r^\prime p^\prime}$. 
Indeed, positive values of $\Phi_\text{IW}$ demonstrate that substantial IW energy radiates outward from the edge of the eddy 
(Fig. \ref{fig:fig2}(a)), as is also visible in the depth-averaged energy flux vector (Fig. \ref{fig:fig2}(b)).
The sign change in $\Phi_\text{IW}$, which occurs at the edge of the eddy, suggests that the spontaneously emitted IWs are generated near the edge of the eddy. Indeed, 
the azimuthal, vertical, and temporal average of the IW flux divergence,
\begin{align}
\label{div_flux}
    \langle \nabla \cdot \bm{F} \rangle_{\theta,z,t} =  \frac{1}{2\pi}
    \frac{1}{H T} \int_0^{2\pi} \int_0^H \int_0^T \frac{\partial}{\partial r} \big(r F_r \big) dt dz d\theta,
\end{align}
is small inside the eddy (the blue shaded region in figure \ref{fig:fig2}(c)), peaks just outside of it, and then decays to zero around $r=30$ km. Further away from the eddy, the value of $\langle \nabla \cdot \bm{F} \rangle_{\theta,z,t}$ remains nearly zero, implying that $\partial/\partial r \langle r F_r \rangle_{\theta,z,t} \approx 0$.  This suggests that the average radial energy flux $\langle F_r \rangle_{\theta,z,t}$ is proportional to $r^{-1}$, consistent with Fig. \ref{fig:fig2}(d).

\section{Spontaneous emission mechanisms}
\label{mechanisms}
Next, we examine the possible processes that can lead to LOB and spontaneous emission\-- namely frontogenesis at the edge of the eddy and eddy instabilities. Geostrophic adjustment \citep{rossby1938mutual} is another obvious candidate for IW emission in an initial value problems. However, because our solutions are statistically steady \citep{barkan2017stimulated} we do not specifically distinguish between geostrophic adjustment and frontogenesis \citep[e.g.,][]{blumen2000inertial}.  
\subsubsection{Frontogenesis}
\label{frontogenessis}
\begin{figure}
    \centering
    \includegraphics[width=\textwidth]{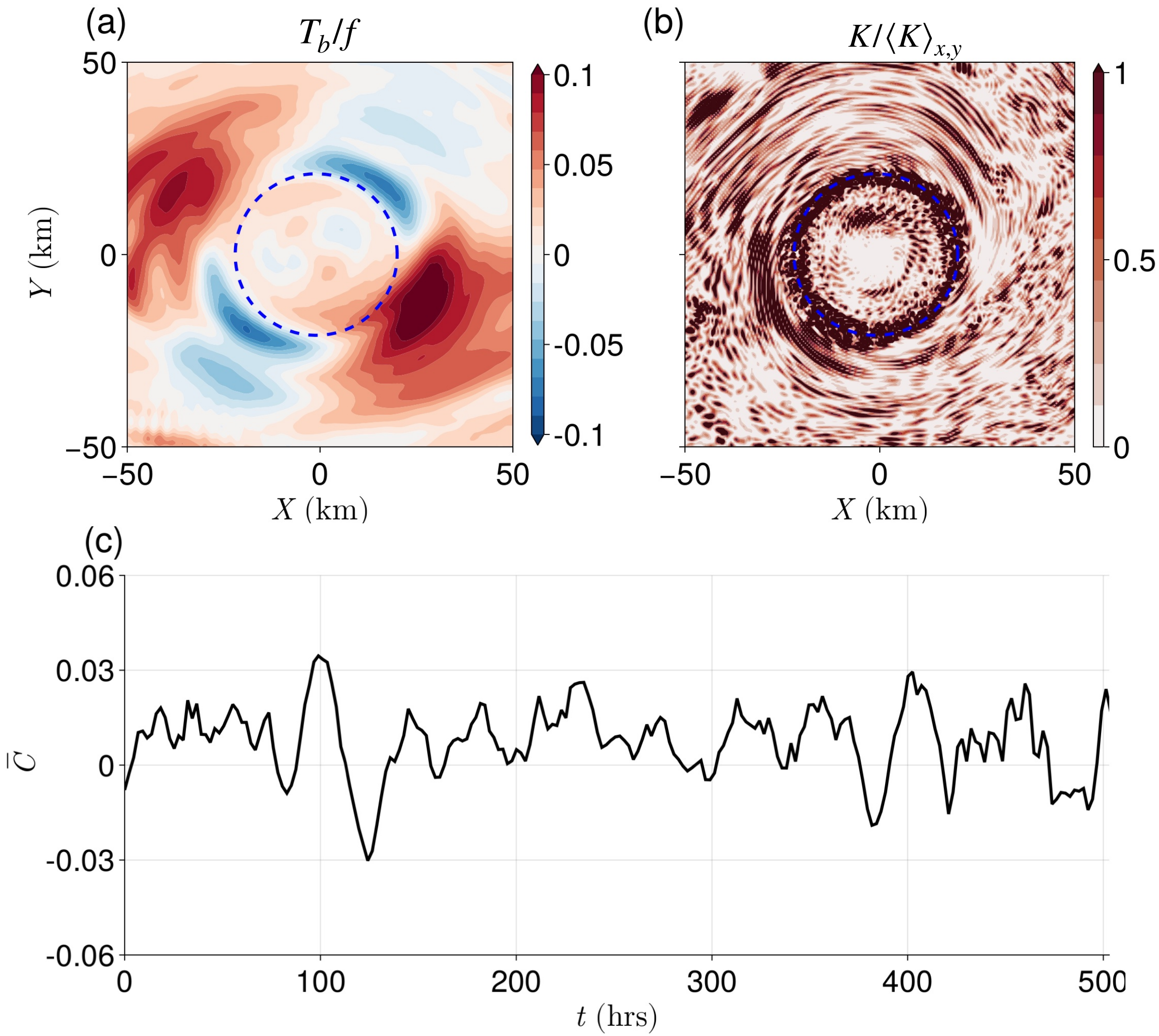}
    \caption{A representative surface snapshot  (a) frontogenetic tendency rate $T_b$ (normalized by $f$) and (b) wave KE $K$ (normalized by the surface average $K$, $\langle K \rangle_{x,y}$). (c) Time series of the correlation function $\overline{C}$ (Eq. \ref{corr_fun}) averaged over the upper $200$m of the domain. The dotted blue lines in panels (a) and (b) mark the edge of the eddy. }
    \label{fig:frontal_tendency}
\end{figure}
To investigate the potential role of frontogenesis in generating IWs, as detailed in \cite{shakespeare2014spontaneous}, we compute the correlation function between the wave kinetic energy (K) and the frontogenetic tendency rate $T_b$ of the mean flow buoyancy gradient, 
\begin{align}
\label{corr_fun}
    \overline{C} = \frac{{\langle T_b K \rangle}_V}{\sqrt{{\langle T_b^2 \rangle}_V {\langle K^2 \rangle}_V}},
\end{align}
where $\langle \,\ \rangle_V$ is the volume integral carried out around the edge of the eddy, i.e., $15 \leq r \leq 25$ km, and over the upper $200$ m of the domain where the strain is substantial (not shown). 
%
%
%
%
In Eq. (\ref{corr_fun}), the frongogenetic tendency rate $T_b$ 
is defined as \citep{barkan2019role} 
\begin{align}
\label{Tb}
    T_b = \frac{\mathcal{F}_b}{|\nabla_h \overline{b}|^2},
\end{align}
with $\mathcal{F}_{b}$ denoting the frontogenetic tendency for  $|\nabla_h \overline{b}^2|$ \citep{hoskins1982mathematical}, 
\begin{align}
    \mathcal{F}_{b} = -\Bigg[\frac{\partial \overline{u}}{\partial X} \Big(\frac{\partial \overline{b}}{\partial X} \Big)^2  + \frac{\partial \overline{v}}{\partial Y} \Big(\frac{\partial \overline{b}}{\partial Y} \Big)^2
    + \Big( \frac{\partial \overline{v}}{\partial X} + \frac{\partial \overline{u}}{\partial Y} \Big) \frac{\partial \overline{b}}{\partial X} \frac{\partial \overline{b}}{\partial Y} \Bigg],
\end{align}
such that positive (negative) values of $T_b$ denote frontogenetic (frontolytic) flow regions. 
The wave kinetic energy is defined as
\begin{align}
\label{wke_def}
    K = \frac{1}{2} (\overline{u^\prime u^\prime} + \overline{v^\prime v^\prime}).
\end{align}
Since $K$ is positive definite by construction, $\overline{C}$ is expected to be positive and close to 1 if frontogenetic regions are strongly correlated with regions of high $K$.

Interestingly, we find the correlation $\overline{C}$ to be slightly positive but very weak (Fig. \ref{fig:frontal_tendency}(c)), suggesting that frontogenesis is unlikely the key mechanism responsible of the observed IW emission. Indeed, a representative snapshot of $T_b$ (Fig. \ref{fig:frontal_tendency}(a)) shows rather weak frontogenetic rates without a clear sign at the eddy periphery, and with little spatial resemblance to the IW kinetic energy patterns (Fig. \ref{fig:frontal_tendency}(b)).

\subsubsection{Eddy Instability} 
\label{instability}
We examine whether the observed LOB in the numerical simulation
is related to an instability of the anticyclonic eddy by examining the necessary criteria for different instabilities. 

\begin{figure}
    \centering
    \includegraphics[width=\textwidth]{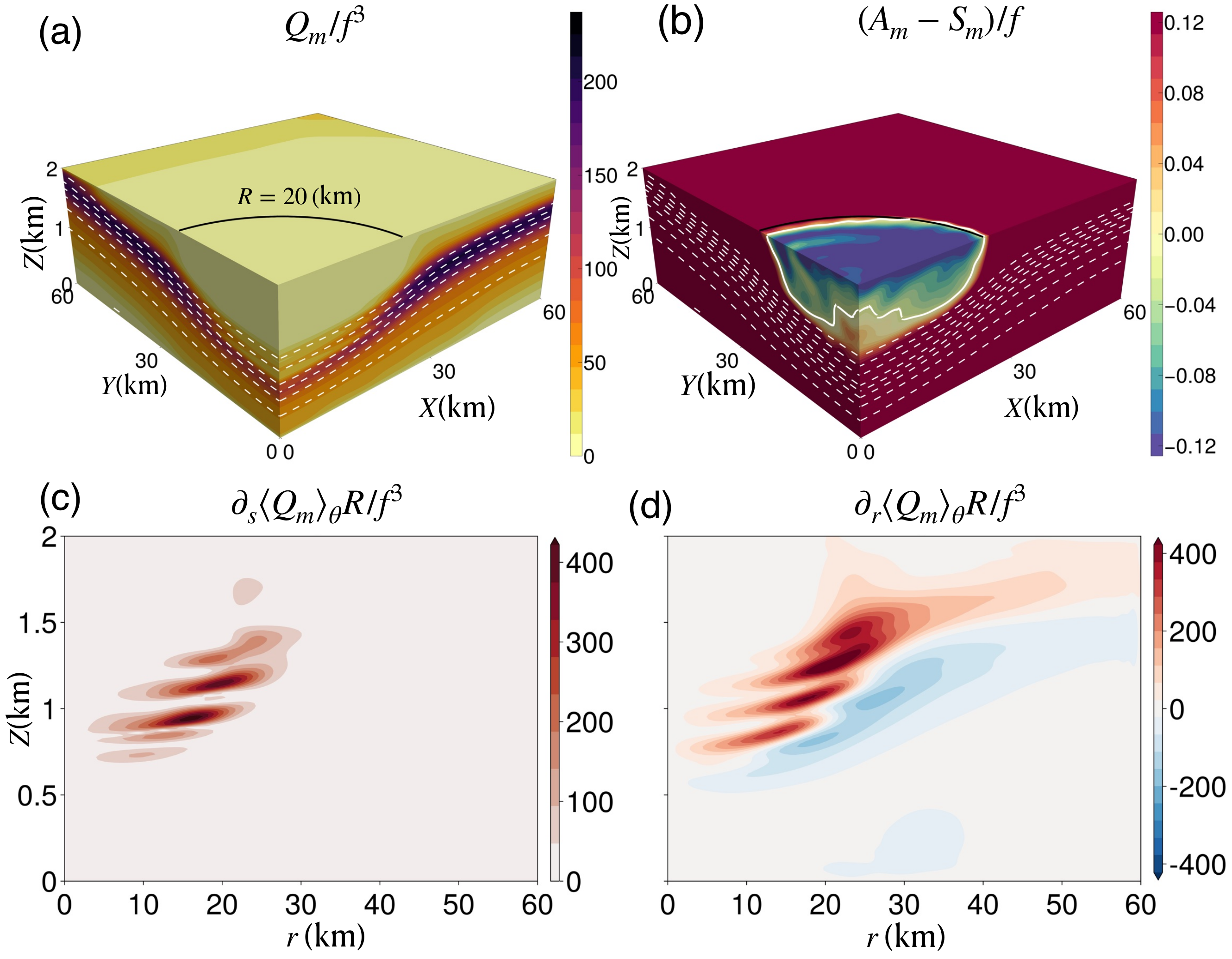}
    \caption{Necessary criteria for instability of the basic state. (a) mean flow PV ${Q}_m$ (normalized by $f^3$; Eq. \ref{def_Q}). (b) $({A}_m-{S}_m)$ (normalized by $f$; Eqs. (\ref{def_A_S})). (c) Along-isopycnal PV gradient $\partial_s \langle {Q}_m \rangle_{\theta}$ (normalized by $f^3/R$; Eq. \ref{isopyc_Q}), and  (d) radial PV gradient $\partial_r \langle \overline{Q}_m \rangle_\theta$ (normalized by $f^3/R$). The white dotted lines in panels (a) and (b) show buoyancy contours with a $0.002$ ms$^{-2}$ contour interval. The solid white lines in panel (b) show where $\overline{A}_m-\overline{S}_m=0$. The black line in panels (a) and (b) mark the edge of the anticyclonic eddy. All quantities are averaged overt $24$ hours. 
    }
    \label{fig:mean_flow}
\end{figure}

Symmetric instability (SI) can trigger LOB and therefore lead to  spontaneous IW emission \citep{chouksey2022gravity}. The necessary condition for symmetric instability requires 
$f{Q}_{m}<0$ \citep{hoskins1974role}, where $f$ is the Coriolis frequency and 
\begin{align}
\label{def_Q}
    {Q}_m = \big(f + {\zeta}_m \big) \partial_Z  {b}_m - \partial_Z  {v}_m \partial_X {b}_m + \partial_Z {u}_m \partial_Y {b}_m 
\end{align}
is the Ertel's PV of the mean flow 
under the hydrostatic approximation \footnote{we solve for the non-hydrostatic equations of motion but because of the grid spacing we use (Section \ref{setup}) our solutions are effectively hydrostatic.} and $\partial_X, \partial_Y, \partial_Z$ denote derivatives in the $X,Y$ and $Z$ directions, respectively.
Since ${Q}_m>0$ in our solutions ($f>0$ in our configuration) the anitcyclonic eddy is stable to SI (Fig. \ref{fig:mean_flow}(a)).

\cite{mcwilliams1998breakdown} and \cite{mcwilliams2004ageostrophic} derived limiting conditions for the integrability of a set of balanced equations in  isopycnal coordinates. They demonstrated that  ${A}_m-{S}_m <0$ (for $f>0$) is a sufficient condition for LOB,
 where
\begin{subequations}
\begin{gather}
\label{def_A_S}
    {A}_m = f + \partial_{X_s} {v}_m 
     - \partial_{Y_s} {u}_m, 
\,\,\,\,\,\text{and    }\,\,\,\,\,
    {S}_m = \sqrt{\Big(\partial_{X_s} {u}_m - \partial_{Y_s} {v}_m \Big)^2 + \Big(\partial_{X_s} {v}_m + \partial_{Y_s} {u}_m \Big)^2},
\tag{\theequation a,b} 
\end{gather}
\end{subequations}
denote the absolute vorticity and the magnitude of the horizontal strain rate of the balanced flow, respectively, and the spatial derivatives are computed 
in the isopycnal coordinate system $(X_s=X, Y_s=Y, Z_s= {b}_m)$
viz.
\begin{subequations}
\begin{gather}
    \frac{\partial}{\partial X_s} = \frac{\partial}{\partial X} 
    - \frac{\partial_X {b}_m}{\partial_Z  {b}_m}
    \frac{\partial}{\partial Z}, 
\,\,\,\,\,\,\,\,\,
    \frac{\partial}{\partial Y_s} = \frac{\partial}{\partial Y} 
    - \frac{\partial_Y {b}_m}{\partial_Z  {b}_m}
    \frac{\partial}{\partial Z}.
\tag{\theequation a,b}    
\end{gather}
\end{subequations}
\cite{menesguen2012ageostrophic} and \cite{wang2014ageostrophic} further showed that this LOB condition is closely related to the onset of ageostrophic anticyclonic instability (AAI), which is triggered  in the neighborhood of, rather than precisely at, $A_m - S_m<0$. 
The simulated anticyclonic eddy in our solutions satisfies this condition for LOB, and may indeed be unstable to AAI (Fig. \ref{fig:mean_flow}(b)).

The necessary condition for an inflection point instability is given by the Rayleigh-Kuo-Fj{\o}rtoft condition, which requires a sign change of the PV gradient within the domain (sometimes referred to as barotropic or lateral shear instability). In the case of a baroclinic flow (as is the case here), the necessary condition is the sign change in the along isopycnal gradient of PV within the domain \citep{eliassen1983charney}, which is defined as
\begin{align}
\label{isopyc_Q}
    \partial_s \langle {Q}_m \rangle_\theta 
     = \partial_r \langle Q_m \rangle_\theta  - 
    \frac{\partial_r \langle {b}_m \rangle_{\theta}}{\partial_z \langle {b}_m \rangle_{\theta}}
    \partial_z \langle {Q}_m \rangle_\theta.
\end{align}
Interestingly, the azimuthal- and time-averaged $\partial_s \langle {Q}_m \rangle_\theta$ does not change sign within the anticyclonic eddy (Fig. \ref{fig:mean_flow}(c)) whereas the azimuthal- and time-averaged $\partial_r \langle {Q}_m \rangle_\theta$ does (Fig. \ref{fig:mean_flow}(d)). This implies that the anticyclonic eddy is stable to inflection point instability but may be unstable to barotropic (lateral shear) instability. Barotropic instability can occur within a balance model (e.g., the QG model) and, therefore, does not necessarily lead to LOB. 
{However, if the Rossby number of the eddy is sufficiently large, the barotropic instability can become radiative. Such radiative instability has been termed Rossby Inertia Buoyancy (RIB) instability \citep[][;see Section 7 for more detail]{schecter2004damping,hodyss2008rossby}. }


Kelvin-Helmholtz instability \citep{miles1963stability}, which can be triggered when the Richardson number $Ri=\partial_z {b}_m/((\partial_z u_m)^2+(\partial_z v_m)^2) <1/4$, can also lead to LOB. However, in our case $Ri>1/4$ everywhere in the domain (not shown). We can further rule out centrifugal instability, which is expected to eventually lead to the breakdown of the anticyclonic eddy over rather rapid time scales \citep{carnevale2011predicting}. Such breakdown is not observed in the numerical simulation (see supplementary movie 1).

%



\section{Linear stability analysis: configuration and numerical methods}
\label{stability_analyis}
In the previous section we showed that the anticyclonic eddy is susceptible to AAI and barotropic shear instability. In this section, we carry out a linear stability analysis of the anticyclonic eddy to determine whether the observed spontaneous IW emission results from an instability. 

\begin{figure}
    \centering
    \includegraphics[width=\textwidth]{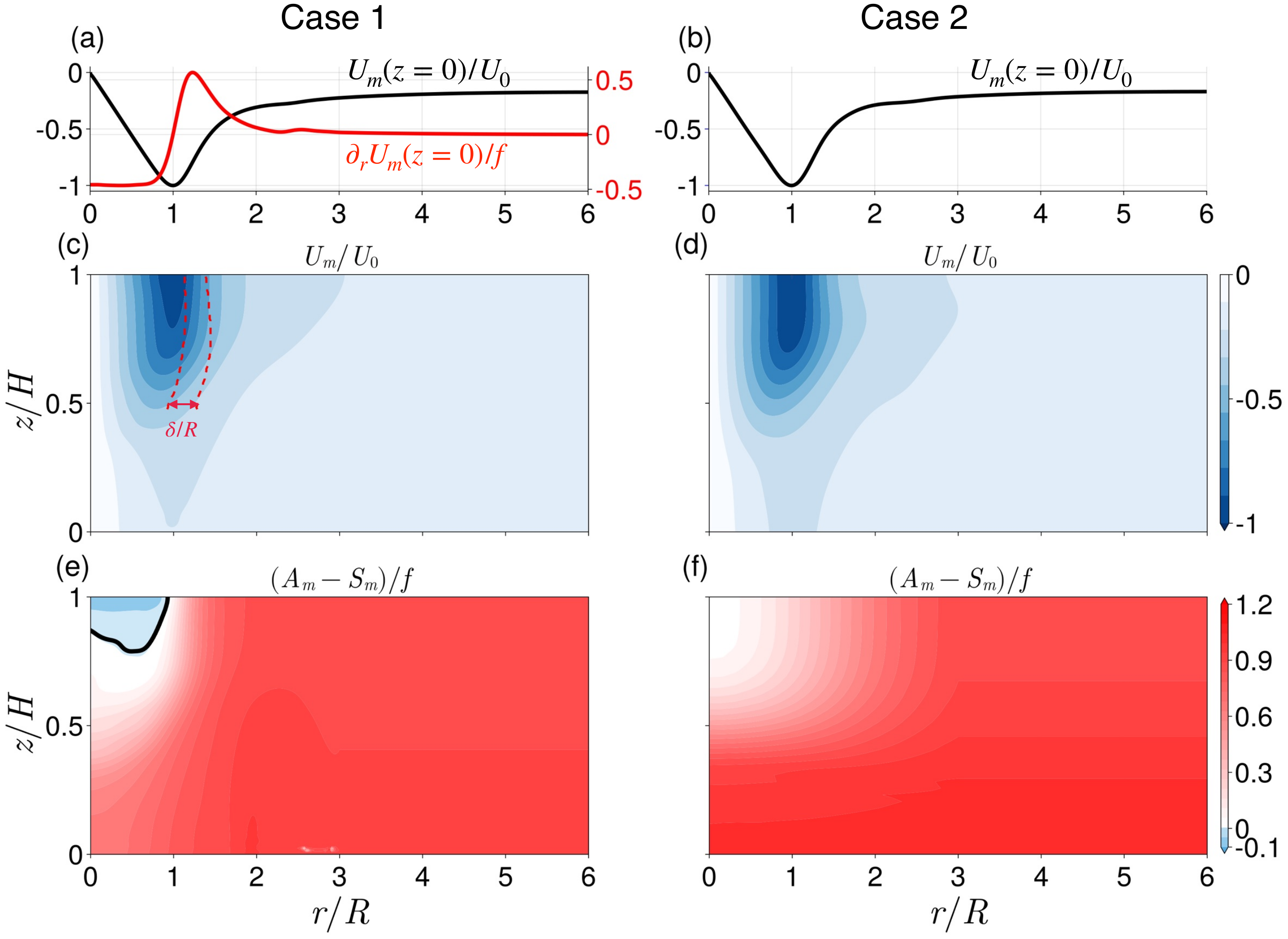}
    \caption{The basic states used for the linear stability analysis. (a,b) Azimuthally-averaged surface azimuthal mean velocity ${U}_m$ (normalized by maximal magnitude of the eddy azimuthal velocity $U_0$), and
    (c,d) contour plots of ${U}_m$ (normalized by $U_0$).
    (e,f) The necessary condition for AAI, where the solid black line in panel (e) denotes the $({A}_m-{S}_m)=0$ contour. Case 1 correspond to a basic state where the necessary condition for AAI is satisfied (e), whereas Case 2 corresponds to a basic state where the necessary condition for AAI is not satisfied (f). The red line in panel (a) shows the surface horizontal shear $\partial_r {U}_m$ (normalized by $f$). The red dotted lines in panel (c) shows the shear layer thickness $\delta$ (normalized by the radius of the eddy $R$) computed based on radial distance corresponding to $80\%$ of the maximum magnitude of $\partial_r {U}_m$.}
    \label{fig:basic_state}
\end{figure}

Our basic state is defined with respect to the azimuthally-averaged and 24-hour low-passed fields (Fig. \ref{fig:basic_state}(a,c,e)), which approximately satisfy gradient wind balance (Fig. \ref{fig:fig3}(b)). This basic state, which we refer to as case 1, satisfies the necessary condition for both AAI and lateral shear instability. 
In what follows, we contrast the stability analysis of the basic state in case 1 with that of a modified basic state (case 2; Fig. \ref{fig:basic_state}(b,d,f)), where we spatially low-pass the normal strain components ($\partial_X {u}_m$ and $\partial_Y {v}_m$) such that $({A}_m-{S}_m) > 0$ everywhere (Fig. \ref{fig:basic_state}(f)). 
The low-pass filter is a sixth-order Butterworth spatial filter with a filter width of $1.5$km.
This comparison allows us to determine which is the dominant instability mechanism that leads to the spontanesous IW emission. 

\subsection{Governing equations}
The equations of motion for the perturbation fields ($u_r,u_{\theta},w,p,b$) satisfy the linearized Navier-Stokes equations on an $f$-plane, under the Boussinesq approximation. We use a cylindrical coordinate system centered around the anticyclonic eddy (Eq. \ref{cylindrical_coordinates}) and define the following length and time scales 
\begin{subequations}
\label{len_scales}
\begin{gather}
    r =  R \tilde{r}, \,\,\,\
    z =  H \tilde{z}, \,\,\,\
    t = \frac{1}{f} \tilde{t}, 
    \tag{\theequation a-c}
\end{gather}
\end{subequations}
where $R=20$ km is the eddy radius, $H=2$ km is the domain depth, and $f=1.2\times 10^{-4}\,\,\text{s}^{-1}$ is the Coriolis frequency used in our simulations. 

The velocity, pressure, and buoyancy are scaled with
\begin{subequations}
\label{flow_scales}
\begin{gather}
\label{flow_scale}
  (u_r, u_\theta) = U_0 (\tilde{u}_r, \tilde{u}_\theta), \,\,\,\
   w = U_0 H/R  \tilde{w}, \,\,\,\
   p = f U_0 R \tilde{p}, \,\,\,\
   b = f U_0 R/H \tilde{b},
   \tag{\theequation a-d}
\end{gather}
\end{subequations} 
where $U_0$ is a characteristic velocity scale, taken to be $1.05$ms$^{-1}$\-- the maximal magnitude of the eddy azimuthal velocity.
Using (\ref{len_scales}a-c) and (\ref{flow_scales}a-d), the 
equations of motion are
\begin{subequations}
\label{nondim_gov_eqs}
\begin{align}
    \frac{D\tilde{u}_r}{D\tilde{t}} -  \Big(1+ 2{Ro} {\widetilde{\Omega}}  \Big) {\tilde{u}_\theta} &= -\frac{\partial \tilde{p}}{\partial \tilde{r}} + 
    {Ek} \Big(\widetilde{{\nabla}}^2 \tilde{u}_r - \frac{1}{\tilde{r}^2} \tilde{u}_r - \frac{2}{\tilde{r}^2} \frac{\partial {\tilde{u}_\theta}}{\partial \theta} \Big),
\\
    \frac{D\tilde{u}_\theta}{D\tilde{t}} +   
    \Big(1 + Ro \tilde{\zeta} \Big) \tilde{u}_r + {Ro} \tilde{r} \frac{\partial \widetilde{\Omega}}{\partial \tilde{z}} \tilde{w} &= -\frac{1}{\tilde{r}}\frac{\partial \tilde{p}}{\partial \theta} + {Ek} \Big(\widetilde{\nabla}^2 \tilde{u}_\theta - \frac{1}{\tilde{r}^2} \tilde{u}_\theta + \frac{2}{\tilde{r}^2} \frac{\partial \tilde{u}_r}{\partial \theta} \Big),
\\
    \frac{D\tilde{w}}{D\tilde{t}} &= -\frac{1}{\alpha^2} \frac{\partial \tilde{p}}{\partial \tilde{z}} + \frac{1}{\alpha^2} \tilde{b} + {Ek} \widetilde{\nabla}^2 \tilde{w}, 
\\
    \frac{D\tilde{b}}{D\tilde{t}} + Ro  \tilde{u} \frac{\partial \widetilde{B}}{\partial \tilde{r}} + {Ro}  \tilde{w} \frac{\partial \widetilde{B}}{\partial \tilde{z}} &=  \frac{Ek}{Pr} \widetilde{\nabla}^2 \tilde{b}, 
\\
    \frac{1}{\tilde{r}}\frac{\partial}{\partial \tilde{r}} (\tilde{r} \tilde{u}_r) + \frac{1}{\tilde{r}} \frac{\partial \tilde{u}_\theta}{\partial \theta} + \frac{\partial \tilde{w}}{\partial \tilde{z}} &= 0,
\end{align}
\end{subequations}
where $\widetilde{U}_\theta$, $\widetilde{\Omega}=\widetilde{U}_\theta/\tilde{r}$, and $\tilde{\zeta}=1/\tilde{r}\partial/\partial \tilde{r}(\tilde{r}^2 \widetilde{\Omega})$ are the nondimensional azimuthal velocity, angular velocity, and vertical component of vorticity of the basic-state, respectively. 
The Rossby number ${Ro} = U_0/(fR)$, and $\alpha=H/R$ is the aspect ratio of the eddy. The Ekman number, ${Ek}=\nu/(fR^2)$, is set to be $10^{-8}$ ({corresponding to a viscosity $\nu=5\times10^{-4}$m$^2$s$^{-1}$, as is used in the numerical simulation}), and the Prandtl number ${Pr}=\nu/\kappa$, is taken to be $1$, where $\kappa$ is the diffusivity. 
The nondimensional material derivative is 
\begin{align}
    \frac{D}{D\tilde{t}} = \frac{\partial}{\partial \tilde{t}} + 
    {Ro} {\widetilde{\Omega}} \frac{\partial}{\partial \theta}, 
\end{align}
and the Laplacian operator is
\begin{align}
    \widetilde{{\nabla}}^2 = \frac{\partial^2}{\partial \tilde{r}^2} + \frac{1}{\tilde{r}} \frac{\partial}{\partial \tilde{r}} + \frac{1}{\tilde{r}^2} \frac{\partial^2}{\partial \theta^2} + \frac{1}{\alpha^2} \frac{\partial^2}{\partial \tilde{z}^2}. 
\end{align}
We consider a normal-mode form of the perturbations 
\begin{equation}
\label{normal_mode}
    [\tilde{u}_r, \tilde{u}_\theta, \tilde{w}, \tilde{p}, \tilde{b}](\tilde{r}, \theta, \tilde{z}, \tilde{t}) =
    \mathfrak{R} \big([\widehat{u}_r, \widehat{u}_\theta, \widehat{w}, \widehat{p}, \widehat{b} ](\tilde{r}, \tilde{z})
    \ee^{\tilde{\omega} \tilde{t} + \ii m \theta} \big), 
\end{equation}
where $\mathfrak{R}$ denotes the real part and the hat quantities denote the complex eigenfunctions, which depend on  $\tilde{r}$ and $\tilde{z}$. The variable $m$ is the azimuthal wavenumber and 
$\tilde{\omega}=\tilde{\omega}_r + \ii \, \tilde{\omega}_i$, with $\tilde{\omega}_r$ denoting the growth rate and $\tilde{\omega}_i$ denoting the frequency of the perturbation. 
In what follows, we consider only the positive $m$ values since $\tilde{\omega}(m) = \tilde{\omega}^\star(-m)$, where the `star' denotes the complex conjugate.
The domain is $\tilde{r} \in [0, \tilde{R}_{max}]$ and $\tilde{z} \in [0, 1]$, {where $\tilde{R}_{max}=9$ is the maximum radial domain size (see section \ref{stability_analyis}\ref{stability_method} and Appendix B for more detail).}


The boundary conditions for the velocity and pressure at $\tilde{r}=0$ depend on the azimuthal wavenumber $m$ \citep{batchelor1962analysis,khorrami1989application}, 
\begin{subequations}
\label{bcs_r0}
\begin{align}
    \frac{\partial \tilde{u}_r}{\partial \tilde{r}} = \tilde{u}_r + \frac{\partial \tilde{u}_\theta}{\partial \theta} = \tilde{w} = \tilde{p} = \tilde{b} = 0, \,\,\,\,\, \text{for} \,\,\ m=1, 
\\
    \tilde{u}_r = \tilde{u}_\theta = \tilde{w} = \tilde{p} = \tilde{b} = 0, \,\,\,\,\, \text{for} \,\,\ m \geq 2. 
\end{align}
\end{subequations}
The boundary conditions at $\tilde{r}=\tilde{R}_{\text{max}}$ are given by
\begin{align}
\label{bcs_r1}
    \tilde{u}_r = \tilde{u}_\theta = \tilde{w} = \tilde{p} = \tilde{b} = 0.
\end{align}
In accordance with the numerical solutions (i.e., \cite{barkan2017stimulated})
 we choose free-slip, rigid wall, and no-flux boundary conditions in the vertical direction, i.e., 
\begin{align}
\label{bcs_z}
    \frac{\partial \tilde{u}_r}{\partial \tilde{z}} = \frac{\partial \tilde{u}_\theta}{\partial \tilde{z}} = \tilde{w} = 
    \frac{\partial \tilde{p}}{\partial z} =
    \frac{\partial \tilde{b}}{\partial z}=0, 
    \,\,\,\,\, \text{at}\,\ \tilde{z}=0, 1.
\end{align}

\subsection{Numerical methodology}
\label{stability_method}
Equations (\ref{nondim_gov_eqs}a-e) are discretized using second-order finite differences. The resulting discretized Eqs. (\ref{nondim_gov_eqs}a-e), using Eq. (\ref{normal_mode}), and with boundary conditions Eqs. (\ref{bcs_r0}a-c), (\ref{bcs_r1}) and (\ref{bcs_z}) can be expressed as a standard generalized eigenvalue problem
\begin{align}
\label{gen_eigvals}
    \bm{\mathcal{A}} \bm{\mathcal{X}} =\tilde{\omega} \bm{\mathcal{B}} \bm{\mathcal{X}},   
\end{align}
where $\tilde{\omega}$ is the eigenvalue, $\bm{\mathcal{X}}=[\widehat{u}_r, \widehat{u}_\theta, \widehat{w}, \widehat{p}, \widehat{b}]^T$ is the eigenvector. The sparse matrices $\bm{\mathcal{A}}$ and $\bm{\mathcal{B}}$ are of size $(5N_rN_z)^2$, with $N_r$ and $N_z$ denoting the number of grid points in the $r$- and $z$-directions, respectively.
The eigenvalue problem in Eq. (\ref{gen_eigvals}) is solved using the FEAST algorithm, which is based on the complex contour integration method \citep{polizzi2009density}. In what follows, we only consider the perturbation mode with the largest growth rate for a given value of $m$. 
The benchmark of the eigensolver is discussed in
Appendix A. 

The grid convergence results (Appendix B) are obtained for the most unstable mode (i.e., $m=7$) by varying the number of grid points from ${N}_z = 50$ to $N_z=100$ while keeping the ratio $N_r/N_z=\tilde{R}_{max}$. Convergence is obtained for $N_z=80$ and $N_r=720$ (Fig. \ref{fig:grid_test}). 
Furthermore, we check the sensitivity of the results to the domain size in the radial direction by comparing between  $\tilde{R}_{max}=6$ and $\tilde{R}_{max}=9$, and find little difference (Fig. \ref{fig:eigs_cmp}). This indicates that our results are not influenced by our choice of boundary conditions. 
In what follows, we present the linear stability results using $N_z=80$, $N_r=720$ and $\tilde{R}_{max}=9$. 

\section{Results of the stability analysis and comparison with the numerical solution}
\label{instab}
\begin{figure}
    \centering
    \includegraphics[width=\textwidth]{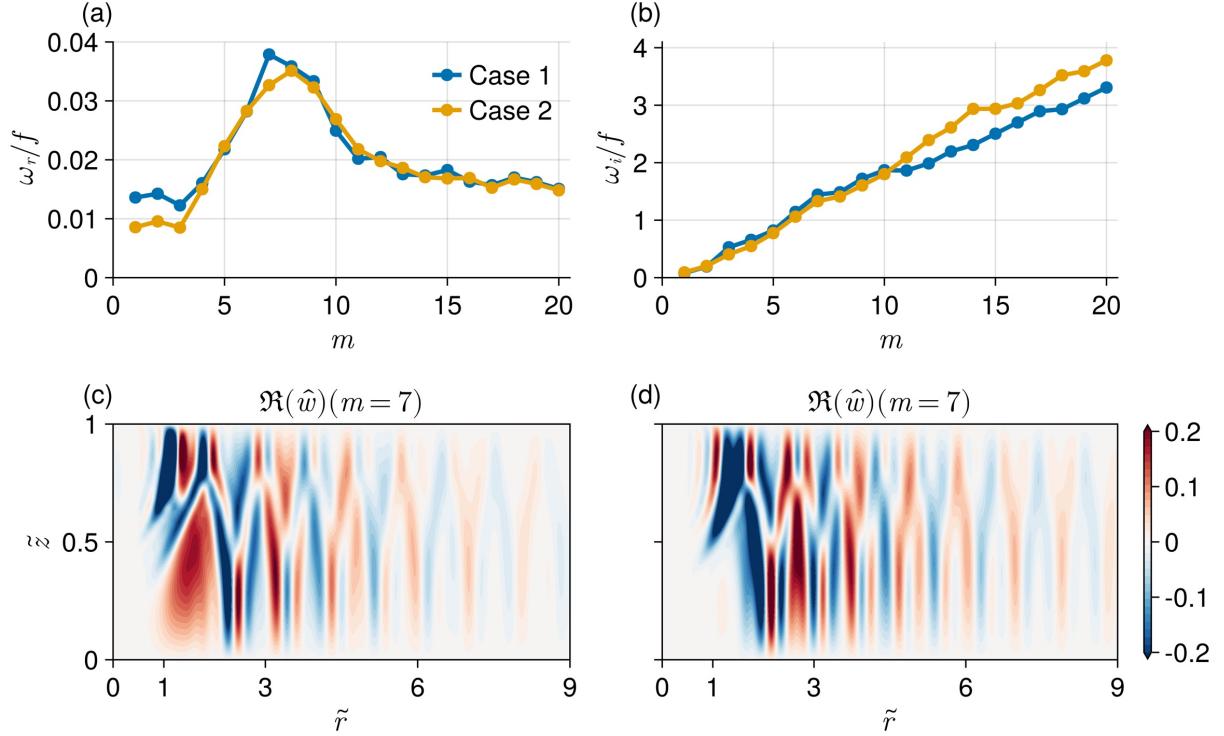}
    \caption{(a) Nondimensional growth rate $\tilde{\omega}_r=\omega_r/f$ and (b) nondimensional frequency $\tilde{\omega}_i=\omega_i/f$ for different values of azimuthal wavenumber $m$, computed for the two basic states (figure \ref{fig:basic_state}). The perturbation frequency $\omega_i$ increases almost linearly with the azimuthal wavenumber $m$. A linear fit of panel (b) data shows that the slope of the curves (i.e., $\tilde{\omega}_i/m$) are $0.17$ and $0.19$ for case 1 and case 2, respectively.
    Panels (c) and (d) show the real part of the vertical velocity eigenfunction $\mathfrak{R}(\hat{w})$ for the two basic states, for $m=7$. 
    }
    \label{fig:growth_rate}
\end{figure}
The linear stability analysis described in the previous section is carried out for the two basic states (Fig. \ref{fig:basic_state}) corresponding to the simulated anticyclonic eddy (case 1) and the smoothed-strain version (case 2; AAI stable).
The growth rates and frequencies for different azimuthal wavenumbers are nearly identical for the two cases (Fig. \ref{fig:growth_rate}(a,b)), with the most unstable modes corresponding to $m=7-9$ (the most unstable mode is $m=7$ and $m=8$ for case 1 and case 2, respectively).  
Furthermore, the eigenfunctions also share similar spatial structures (Figs. \ref{fig:growth_rate}(c,d)), with a clear signature of a radiating IW that closely resembles the spiral shaped IWs emanating from the edge of the eddy in the numerical solution (Figs. \ref{fig:fig1}(c,d)). Although it is possible that some weakly unstable AAI modes are also excited in case 1 (we only look for the most unstable modes in our analysis), these findings suggest that the spontaneous IW emission in the numerical solution is likely result of a radiative instability. 

\subsection{Kinetic energy exchanges}
\label{ke_exchange}
To further establish the connection between the linear stability analysis and the numerical solution we compare the exchange terms in the evolution equation of perturbation KE. 
Due to a near axisymmetric structure of the eddy (e.g., Fig. \ref{fig:fig1}(a)), it is reasonable to define the perturbation quantities in the numerical simulation as the deviation from the azimuthal average. 
With this definition, the dominant energy exchange terms can be expressed as \footnote{the radial and vertical components of the mean flow are negligible compared with the azimuthal component}
\begin{subequations}
\begin{gather}
\label{perturb_eqs_model}
	\text{HSP} = -{u}^\prime_r {u}^\prime_\theta \frac{\partial \langle {U_\theta} \rangle_{\theta}}{\partial r}, 
\,\,\,\,\,\
	\text{VSP} = -{w}^\prime {u}^\prime_\theta  
    \frac{\partial \langle {U_\theta} \rangle_{\theta}}{\partial z}, 
\,\,\,\,\,\
	\text{BFLUX} = {w}^\prime {b}^\prime,
    \tag{\theequation a-c}
\end{gather} 
\end{subequations}
where $\langle {U_\theta} \rangle_{\theta}$ is the azimuthally-averaged azimuthal velocity of the eddy, and the primes denote perturbations from the azimuthal-mean. We verified that the perturbation quantities are an order of magnitude smaller than the maximal magnitude of the azimuthal velocity, consistent with linear stability theory.
The first two terms in Eq. (\ref{perturb_eqs_model}), horizontal shear production (HSP) and vertical shear production (VSP), are associated with the horizontal (radial) and vertical shear of the mean flow, respectively. A positive value of HSP (or VSP) describes the growth of the perturbation KE at the expense of the mean flow KE. The third term in Eq. (\ref{perturb_eqs_model}), the buoyancy flux (BFLUX), quantifies energy exchanges between perturbation kinetic and potential energies.

\begin{figure}
    \includegraphics[width=\textwidth]{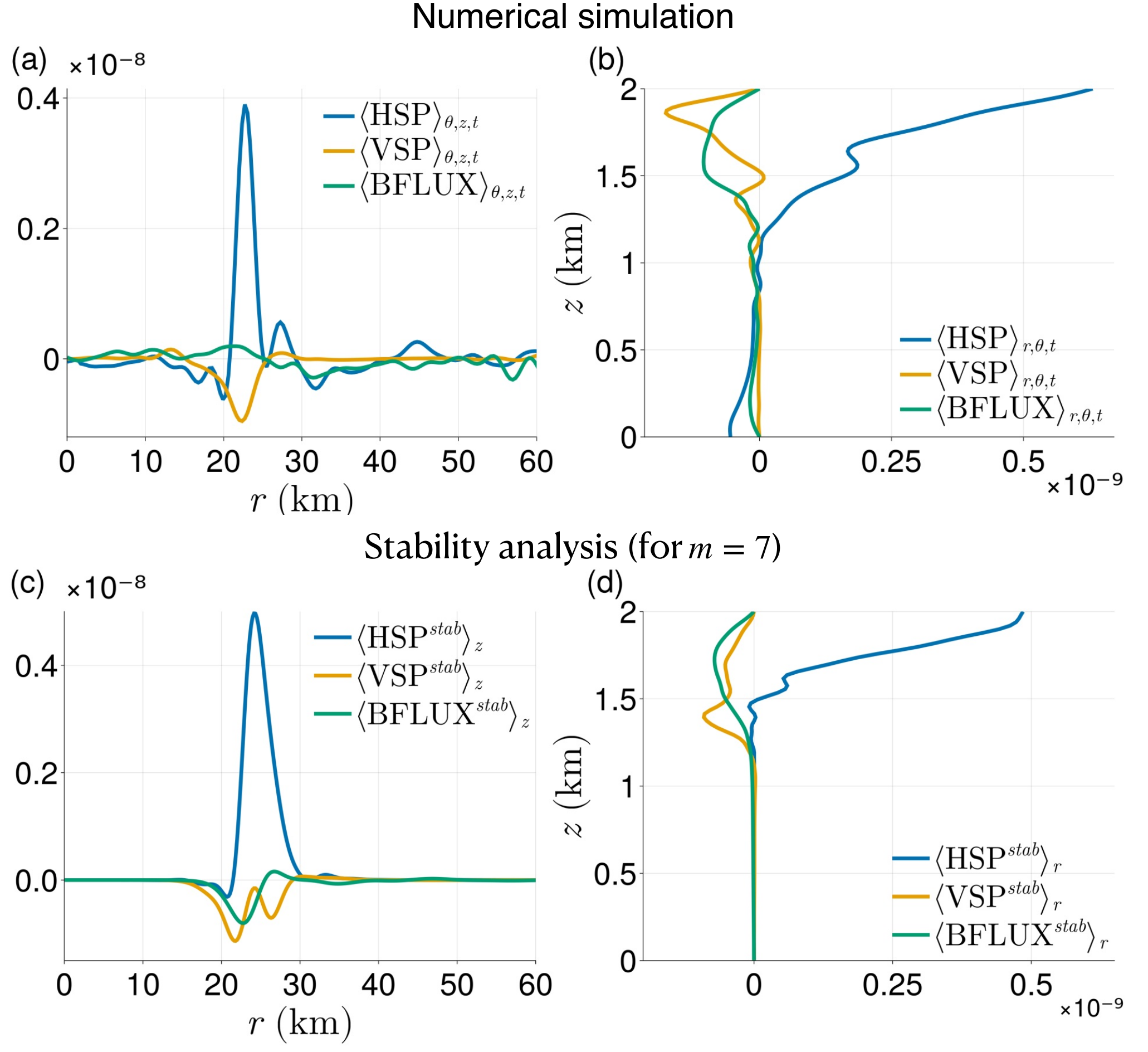}
    \caption{A comparison of the energy exchange terms between the mean flow and the perturbation, computed in the numerical simulations (panels (a,b); Eqs. (\ref{perturb_eqs_model})) and in the stability analysis of the case 1 with $m=7$ (panels (c,d); Eq. (\ref{pertb_ke}); superscript \textit{stab}). The horizontal shear production (HSP), vertical shear production (VSP), and the buoyancy flux (BFLUX) are averaged over depth, azimuth and time in (a) and over radius, azimuth, and time in (b). The time average in panels (a,b) is over $24$ hours. 
    Similarly, HSP$^\text{stab}$, VSP$^\text{stab}$, and BFLUX$^\text{stab}$ are depth-averaged and radially averaged in panels (c) and (d), respectively.
 The terms HSP$^\text{stab}$, VSP$^\text{stab}$, and BFLUX$^\text{stab}$ are dimensionalized using Eqs. (\ref{flow_scale}). The perturbation quantities in the stability analysis are multiplied with a constant, which is obtained by matching the maximal magnitude of $w$ from the stability analysis with the maximal magnitude of $w^\prime$ at the radial location where HSP peaks (panel (a)).
 All quantities are expressed in units of W kg$^{-1}$.} 
    \label{fig:pertb1}
\end{figure}
The following perturbation KE equation \-- corresponding to the linear stability analysis \-- is obtained  
by substituting Eq. (\ref{normal_mode}) into 
Eqs. (\ref{nondim_gov_eqs}), and multiplying Eqs. (\ref{nondim_gov_eqs}a), (\ref{nondim_gov_eqs}b) and (\ref{nondim_gov_eqs}c),
with $\widehat{u}^\star$, $\widehat{v}^\star$ and $\widehat{w}^\star$, respectively,
\begin{align}
\label{pertb_ke}
    2 \tilde{\omega} \Big \langle K_p \Big \rangle_\theta 
    + \underbrace{Ro\frac{\widetilde{U}}{\tilde{r}} \Big \langle\widehat{u_r} \widehat{u_\theta}^\star - 2\widehat{u_r}^\star \widehat{u_\theta} \Big \rangle_\theta}_{\widetilde{\text{Curvature}}}
    + \underbrace{\Big \langle \widehat{u_r} \widehat{u_\theta}^\star - \widehat{u_r}^\star \widehat{u_\theta} \Big \rangle_\theta}_{\widetilde{\text{Coriolis}}} 
    = \underbrace{-Ro \frac{\partial \widetilde{U}}{\partial \tilde{r}} \Big \langle \widehat{u_r} \widehat{u_\theta}^\star \Big \rangle_\theta}_{{\widetilde{\text{HSP}}}^{stab}}
     \underbrace{-Ro \frac{\partial \widetilde{U}}{\partial \tilde{z}} \Big \langle \widehat{w} \widehat{u_\theta}^\star \Big \rangle_\theta}_{{\widetilde{\text{VSP}}}^{stab}}
     \nonumber \\
     + \underbrace{\Big \langle \widehat{w}^\star \widehat{b} \Big \rangle_\theta}_{{\widetilde{\text{BFLUX}}}^{stab}}
     + \underbrace{\widetilde{\nabla} \cdot  \Big \langle\widehat{\bm{u}}^\star \widehat{p} \Big \rangle_\theta}_{\widetilde{\text{PWORK}}}
     + \underbrace{Ek \Big \langle \widehat{u_r}^\star \widetilde{\nabla}^2 \widehat{u_r} 
     - \frac{1}{\tilde{r}^2} \widehat{u_r} \widehat{u_r}^\star + \widehat{u_\theta}^\star \widetilde{\nabla}^2 \widehat{u_\theta} 
     - \frac{1}{\tilde{r}^2} \widehat{u_\theta} \widehat{u_\theta}^\star
     + \alpha^2 \widehat{w}^\star \widetilde{\nabla}^2 \widehat{w} \Big \rangle_\theta}_{\widetilde{\text{DISP}}}
\end{align}
where  $K_p=1/2(\widehat{u_r}\widehat{u_r}^\star + \widehat{u_\theta}\widehat{u_\theta}^\star + \alpha^2 \widehat{w}\widehat{w}^\star)$ is the perturbation KE. The superscript `stab' is added to the HSP, VSP, and BFLUX to distinguish them from the exchange terms defined in the numerical solution (Eq. (\ref{perturb_eqs_model})), but their physical interpretation remains the same.
The {Curvature} term appears due to the circular structure of the mean flow. 
It is purely imaginary and thus does not contribute to the growth of the perturbation KE. Similarly, the {Coriolis} term does not participate in the growth of the perturbation KE either. 
 The PWORK term denotes KE propagation due to pressure perturbations. It has a zero domain average because there is no KE propagation through the boundaries. 
The dissipation term (DISP) for the unstable perturbation is negligible (not shown). 

{The comparison between the energy exchange terms in the numerical solution and the linear stability analysis for case 1 shows a reasonable agreement (Fig. \ref{fig:pertb1}).
To obtain the magnitude of the energy exchange term in the stability analysis we multiply the perturbation fields 
  $\hat{u_r},\hat{u_\theta}, \hat{w},$ and $\hat{b}$ by a constant 
  that is defined such that $|\hat{w}|=|w'|$ at the radial location where HSP peaks.}
The dominant KE energy exchange term is the HSP (Eqs. (\ref{perturb_eqs_model}) and (\ref{pertb_ke})), which is characteristic of lateral shear instability. 
The radial distributions of $\langle \text{HSP} \rangle_{\theta,z,t}$ and $\langle \text{HSP}^{stab} \rangle_{z}$ show that the energy exchange occurs just outside of the anticyclonic eddy (Fig. \ref{fig:pertb1}(a,c)), where the horizontal shear of the mean flow is positive (e.g., red line in Fig. \ref{fig:basic_state}(a)). This is due to the perturbation phase lines being tilted against the horizontal shear of the mean flow.
The vertical distributions of $\langle \text{HSP} \rangle_{\theta,z,t}$ and $\langle \text{HSP}^{stab} \rangle_{z}$ suggest that the energy exchange occurs in the upper half of the domain (Fig. \ref{fig:pertb1}(b,d)).

\cite{menesguen2012ageostrophic} performed linear stability analysis of an idealized AAI unstable basic state and showed that the AAI growing modes had equal contributions from both HSP and VSP. Since VSP is negligible in our solution (orange lines in Fig. \ref{fig:pertb1}) and because similar dominant energy exchange terms are found for case 2 (not shown), it is unlikely that the MOST unstable modes in our solution are associated with AAI.

\subsection{Phase speed}
\label{phase_speed}
Next, we evaluate whether the radial phase speed $c_p$ predicted by the linear stability analysis agrees with the computed phase speed of the spontaneously emitted IWs in the numerical solution. 
By definition, 
\begin{equation}
\label{def_cp}
    c_p=\omega_i/k_h,
\end{equation}
where $\omega_i$ is the frequency, and $k_h=\sqrt{k^2+l^2}$ is the horizontal (radial) wavenumber, with $k$ and $l$ denoting the $x$ and $y$ wavenumber components, respectively.  

\begin{figure}
    \centering
    \includegraphics[width=\textwidth]{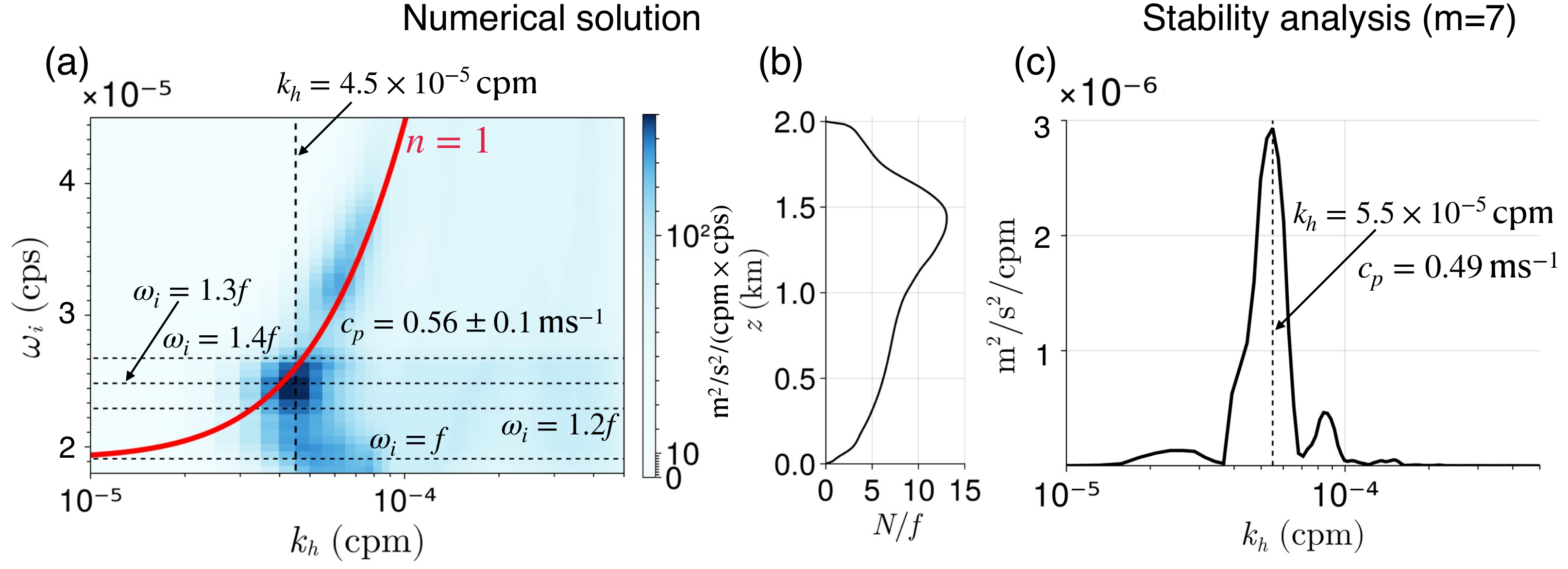}
    \caption{(a) Frequency-horizontal wavenumber power spectral density for the modeled vertical velocity $w$, at $z=1.5$km. 
    The solid red line represents the theoretical estimate of the dispersion relation using Eq. (\ref{freq_esti}) for vertical mode $n=1$. The horizontal dashed black lines mark the frequencies 
    $\omega=f, 1.2f, 1.3f$ and $1.4f$.
    (b) The time- and horizontally averaged normalized stratification profile $N/f$, computed in the red box displayed in figure \ref{fig:fig1}(c), excluding the anticyclonic eddy region (time average is carried out over $35$ inertial periods).
    (c) The horizontal wavenumber power spectral density of the vertical velocity $w$ (using Eq. (\ref{normal_mode}) at $t=0$ after dimensionalize) at $z=1.5$km,  based on the linear stability analysis of case 1, with $m=7$. 
    The power spectra density in panel (a) peaks in the range $1.2f < \omega_i (\text{cps}) < 1.3f$ and $4 \times 10^{-5} < k_h (\text{cpm}) < 5 \times 10^{-5}$, yielding a phase speed estimate of $c_p=0.56 \pm 0.1$ ms$^{-1}$ (Eq. (\ref{def_cp})).
    The horizontal wavenumber (panel (c)) and radial phase speed predicted by the stability analysis are $k_h=0.49$(\text{cpm}) and $c_p=0.49\text{m} \text{s}^{-1}$, using $\omega_i=1.42f$ ($m=7$ in Fig. \ref{fig:growth_rate}(b)).}
    \label{fig:calc_cp_fig}
\end{figure}

In the numerical solution, $c_p$ is computed by fitting dispersion curves to the frequency-horizontal wavenumber power spectral density of the modeled vertical velocity (Fig. \ref{fig:calc_cp_fig}(a). This is done 
by solving a Sturm-Liouville boundary value problem for the IW vertical modes \citep{gill1982atmosphere}, 
\begin{align}
\label{eigs_gill}
    \frac{\partial}{\partial z} \Big( \frac{f^2}{N^2} \frac{\partial \mathscr{F}_n}{\partial z} \Big) = -\frac{1}{\mathrm{R}_n^2} \mathscr{F}_n,
\end{align}
where $\mathscr{F}_n$ denotes the eigenfunction and $\mathrm{R}_n$ denotes the deformation radius for the $n$th vertical mode, and subject to the boundary conditions
$\partial_z \mathscr{F}_n=0$ at $z=0,H$. 
The resulting IW dispersion relation (red line in Fig. \ref{fig:calc_cp_fig}a), computed from 
\begin{align}
\label{freq_esti}
    \omega_i = f \sqrt{1+\mathrm{R}_n^2 k_h^2}
\end{align}
using the time- and horizontally-averaged (excluding the eddy region) buoyancy frequency $N$ (Fig. \ref{fig:calc_cp_fig}(b)),  shows a good agreement with the modeled power spectral density. 


In the linear stability analysis, the frequency $\omega_i$ is directly computed for the various unstable modes (Fig. \ref{fig:growth_rate}(b)). The corresponding horizontal wavenumbers are estimated by computing the horizontal-wavenumber power spectral density of the vertical velocity $w$ for a given mode $m$  (Fig. \ref{fig:calc_cp_fig}(c)).

The resulting $k_h$ and associated $c_p$ (Eq. (\ref{def_cp})) are well within the range of the numerically computed phase-speed (Fig. \ref{fig:calc_cp_fig}(a) and (c)), supporting the premise that the spontaneously emitted IWs result from a radiative instability of the antiyclonic eddy. 



\section{Discussion}
\label{discussion}
\begin{figure}
\includegraphics[width=\textwidth]{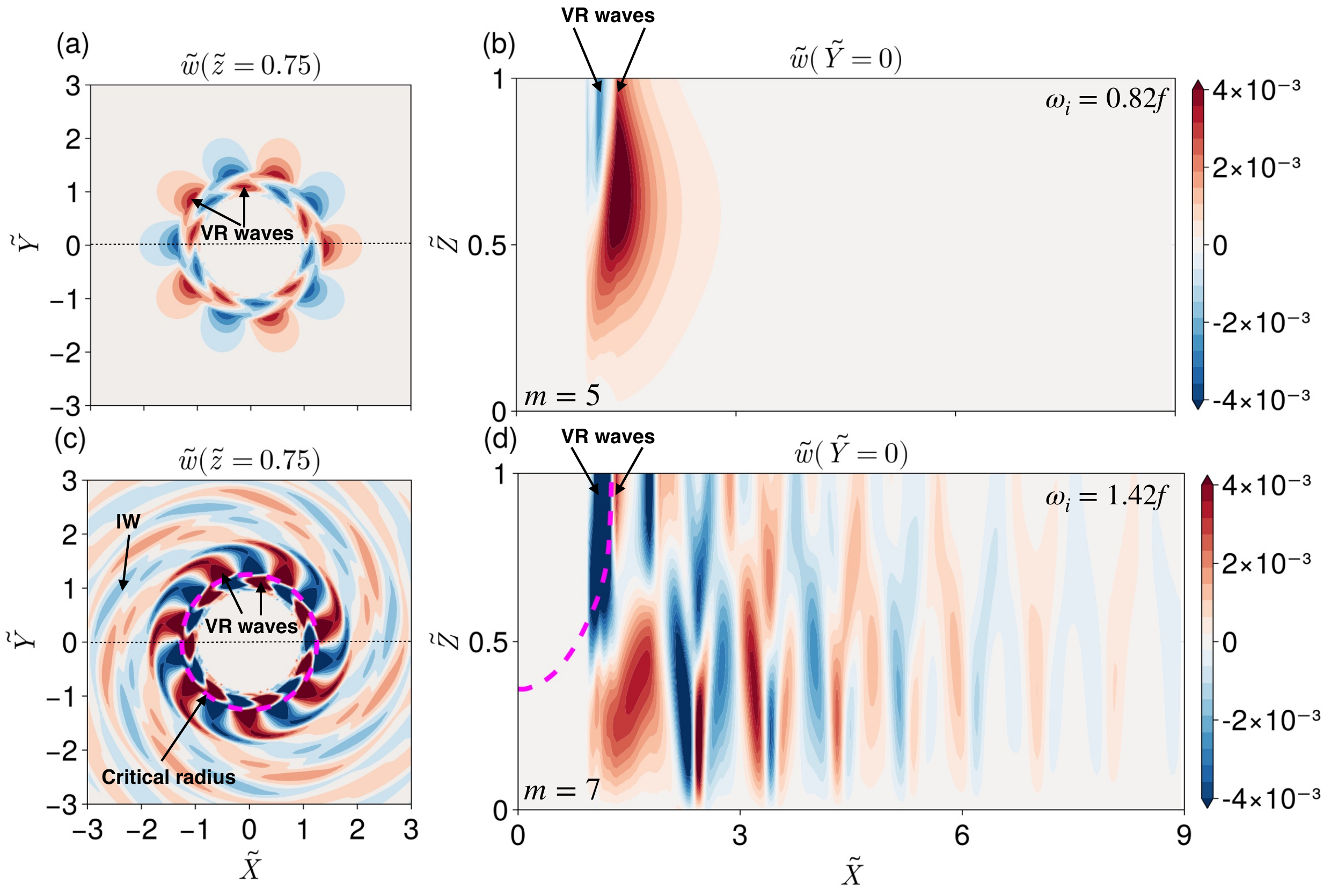}
    \caption{Full solution of the perturbation vertical velocity $\tilde{w}$ at $\tilde{z}=0.75$ for the case 1 from the linear stability analysis is constructed using Eq. (\ref{normal_mode}) at $t=0$ for panels (a,b) $m=5$ and panels (c,d) $m=7$. $(\tilde{X}, \tilde{Y})=(X/R,Y/R)$. 
    The dashed magenta lines in panels (c,d) indicates the critical radius $\tilde{r}_c(\tilde{z}=0.75)$ where $Ro \widetilde{\Omega} = -\tilde{\omega}_i/m$ 
    (nondimensioanl form of Eq. (\ref{critical_radii})). 
    The perturbation frequency $\omega_i$ of $m=5$ and $m=7$ are marked at the top corner of the panels (b) and (d), respectively. The thin black lines in panels (a,c) shows $\tilde{Y}=0$.
    For $m=5$, the perturbation frequency $\omega_i$ is a subinertial frequency; thus, there is no radiative IW. Conversely, for $m=7$, the perturbation frequency $\omega_i$ is a superinertial frequency leading to the spiral shaped radiative IW from the eddy. 
    }
    \label{fig:w_m6_m7}
\end{figure}

{The spontaneous radiation of IW from the eddy in the numerical simulation, can be understood following the RIB instability mechanism discussed in \citet[][hereinafter SM04]{schecter2004damping}. }
In the classical barotropic instability \citep[e.g.,][]{hoskins1985use}, the mechanism leading to perturbation growth can be rationalized as the phase-locking of two counter-propagating 
vortex Rossby waves (VRWs),\footnote{VRWs are analogous to  planetary Rossby waves that propagate on meridional PV gradients \citep{montgomery1997theory}. The term first appeared in the context of atmospheric hurricanes \citep{macdonald1968evidence}.}
located in regions of opposite signs of the radial (horizontal) PV gradient.
In contrast, the RIB instability mechanism described by SM04 relies on an interaction between the exterior VRW and an outward propagating IW. Using linear perturbation theory of an a cyclonic Rankine vortex, they showed that the deformation of the vortex PV surface triggers a VRW with frequency $\omega_i$. When $|\omega_i|>f$, the VRW excites an outward propagating IW with the same frequency. 
This radiative instability relies on the existence of a critical layer, where the angular VRW phase velocity $\omega_i/m$ matches with the angular velocity of the eddy $\Omega$. 
The location of the critical layer is then defined by the resonance condition
\begin{subequations}
\label{critical_radii}
\begin{gather}
    \Omega(r_c) = -\omega_i/m, \,\,\,\,\,\,\,\,\,\
    \Rightarrow \,\,\,\, Ro_l(r_c) = -\frac{1}{m}\frac{\omega_i}{f}, 
    \tag{\theequation a-b}
\end{gather}
\end{subequations}
where $Ro_l=\Omega/f$ is the local Rossby number of the eddy. 
\cite{hodyss2008rossby} and \cite{park2012radiative} extended the work of SM04 and showed the prevalence of this radiative instability in a baroclinic cyclonic eddy and in a barotropic anticyclonic eddy, respectively. In the former case, the perturbation growth rate was found to be somewhat reduced compared with the barotropic case. 
In this article, we demonstrate for the first time the emergence of this radiative instability in forced-dissipative solutions of the Boussinesq equation of motion. For illustration purposes, we contrast the eigenmode structures of two unstable modes (Fig. \ref{fig:w_m6_m7}): $m=5$ - corresponding to a subinertial perturbation frequency ($\omega_i=0.82f$; 
Fig. \ref{fig:growth_rate}(a)), and $m=7$ - the most unstable mode corresponding to a superinertial perturbation frequency ($\omega_i=1.42f$; Fig. \ref{fig:growth_rate}(a)).

For $m=5$ (Figs. \ref{fig:w_m6_m7}(a,b)), the eigenmode structure shows two radial maxima, corresponding to two counter-propagating VRWs, and no IW signature. Conversely, for $m=7$ (Figs. \ref{fig:w_m6_m7}(c,d)), a distinct spiral pattern of IW is visible (consistent with the numerical solution; Fig. \ref{fig:fig1}(c)) that radiates out from the exterior VRWs situated at the critical layer predicated by the SM04 mechanism (Eq. (\ref{critical_radii})).
Similar to $m=5$, there are still two counter-propagating VRWs that can induce mutual amplification through phase locking. However now, the amplification of the exterior VRW can further enhance the interaction with the outward propagating IW, thereby making the spontaneous IW emission a self-sustained process. 

To estimate the magnitude of $Ro_l$ at the vicinity of the critical layer in our solution we consider a shear layer of thickness $\delta$, defined based on the radial distance corresponding to 80\% of the maximal radial shear magnitude at every depth (only the top half of the domain is considered; red dotted line in Fig. \ref{fig:basic_state}(c)). The associated depth averaged azimuthal velocity gives $|\Ro_l|\approx 0.19$. This value is consistent with the observed transition from non-radiating to radiating instability occurring around $m=5-6$ (Fig. \ref{fig:growth_rate}(b)).


Finally, we note that both the structure of the eigenfunctions and the estimated $|Ro_l|$  are very similar for case 2 (not shown). 
This lends further support to the interpretation of the observed insatiability as a radiative instability, following the mechanism proposed by SM04.

\section{Summary}
\label{summary}
In this study we investigate in detail the processes leading to spontaneous IW emission from an anticyclonic eddy in the $O(1)$ Rossby number regime.  
We utilize a high-resolution, forced-dissipative channel solution of the Boussinesq equations of motion and show that spontaneous loss of balance (LOB) around the edge of the eddy closely coincides with the location of IW emission. Furthermore, we carry out perturbation KE analysis and 2D linear stability analysis of the eddy and demonstrate that the LOB and subsequent spontaneous emission occurs due to a radiative instability, following the mechanism proposed by \cite{schecter2004damping}.
To our knowledge, this is the first demonstration of this radiative mechanism in a forced-dissipative Boussinesq solution.
In contrast with centrifugal instability \citep{carnevale2011predicting} and ageostrophic anticyclonic instability \citep{mcwilliams1998breakdown,menesguen2012ageostrophic}, this radiative instability is not specific to anticyclonic eddies and can occur in cyclonic eddies as well, provided they are in the $O(1)$ Rossby number regime. 
In our idealized, high-latitude, channel solution, the spontaneous emission results in a
time-averaged IW energy flux of $0.2 \text{mW}/\text{m}^2$, which is somewhat weaker than the values reported by \cite{alford2013observations}, for a subtropical frontal jet. Nevertheless, if ubiquitous, this radiative instability mechanism can still provide a non-negligible source of IW energy. 

To identify this mechanism in oceanic observations, it is necessary to collect measurements of the velocity field along an eddy cross section \citep[e.g.,][]{l2023ocean}. 
This will allow to estimate the radial shear of the azimuthal velocity $\partial \Omega/\partial r$, from which the shear layer thickness, $\delta$, and the local Rossby number $\Omega/f$ can be estimated (e.g., Fig. \ref{fig:basic_state}(c)). 
According to our stability analysis, the azimuthal wavelength of the most unstable mode is approximately $2 \delta$, which gives an azimuthal wavenumber $m \approx \pi R/\delta$. Thus, the instability can be of radiative type if $(\pi R/\delta )|\Omega|/f > 1$.


In our analysis we ignored the eddy ellipticity, which has previously been shown to affect the stability characteristics under some circumstances \citep{ford1994response,plougonven2002internal}. In addition, we have not examined the pathways of the spontaneously emitted IWs towards dissipation and mixing, either through non-linear wave-wave interactions \citep[e.g.,][]{mccomas1977resonant} or wave-mean flow interactions \citep[e.g.,][]{shakespeare2015spontaneous, nagai2015spontaneous}. Such endeavors are left for future work.

\clearpage
\acknowledgments
SK and RB were supported by ISF grants 1736/18 and 2054/23. The authors report no conflicts of interest.

\datastatement
The linear stability code used in this study is available 
at \url{https://github.com/subhk/Radiative_Shear_Instability}.

\appendix[A]
\label{aap_a}
\appendixtitle{Benchmark of the linear stability code}
\label{benchmark_code}
The stability code used in this study is benchmarked using the results 
of \cite{yim2016stability}. \cite{yim2016stability} carried out a linear stability analysis of an axisymmetric eddy with azimuthal velocity $U$ of the form  
\begin{align}
    U(r,z) \equiv r \Omega(r,z) = r \Omega_0 \ee^{-r^2/R^2-z^2/H^2},
\end{align}
where $R$ is radius of the eddy, $H$ is its half-thickness, and $\Omega_0$ is the maximum value of its angular velocity $\Omega$. The basic state is in gradient wind balance \citep{holton1973introduction}, i.e., 
\begin{align}
    \Big(\frac{2U}{r} + f \Big) \frac{\partial U}{\partial z} = \frac{\partial B}{\partial r},
\end{align}
with
\begin{align}
    B(r, z) = \overline{B}(z) + \alpha^2 (\Omega + f) \Omega.
\end{align}
$\overline{B}(z)=N^2 z$, the buoyancy frequency $N$ is a positive constant, and $\alpha=H/R$.  
The characteristics velocity scale is $U_0 = |\Omega_0| R$ and the Rossby number ${\text{Ro}}=\Omega_0/f$ {\footnote{In \cite{yim2016stability}, ${{Ro}}$ is defined as
${{Ro}}=2\Omega_0/f$.}}. 
The Reynolds number Re is defined as ${Re} = (\Omega_0 R^2)/\nu = {Ro}/{Ek}$, where the Ekman number ${Ek}=\nu/(fR^2)$, and the Froude number is defined as ${Fr}=|\Omega_0|/N$.  
The domain size is take to be $[0, 10R]$ and $[-5H, 5H]$. 
The perturbation boundary conditions at $r=0$ and $r=R$ are similar to Eqs. (\ref{bcs_r0}a-b) and Eq. (\ref{bcs_r1}), respectively. The boundary condition in the vertical direction,  
\begin{align}
    {u}_r = {u}_\theta = {w} = {p} = {b}=0, 
    \,\,\,\,\, \text{at}\,\ z=-5H, 5H.
\end{align}
The number of radial and vertical grid points are $N_r=200$ and $N_z=200$, respectively.

\begin{figure}
    \centering
    \includegraphics[width=\textwidth]{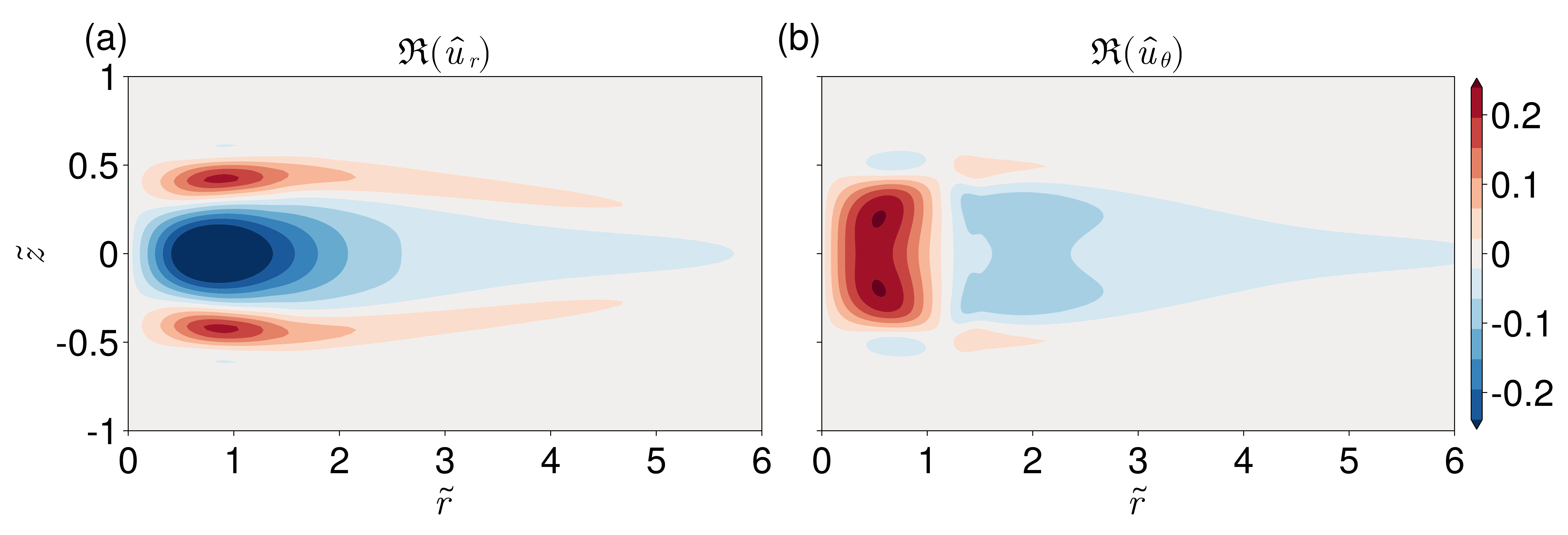}
    \caption{The real part of (a) the radial velocity eigenfunction $\mathfrak{R}(\hat{u}_r)$, and (b) the azimuthal velocity eigenfunction $\mathfrak{R}(\hat{u}_\theta)$ for the most unstable mode ($m=2$), with $\text{Ro}=10$, $\alpha=1.2$, $\text{Fr}=0.5$ and $\text{Re}=10^4$. These results compare well with Fig. 13(a) in \cite{yim2016stability}.}
    \label{fig:benchmark}
\end{figure}

A comparison of the maximum growth rates of the perturbations for different parameters are listed in Table \ref{tab:appendix_tab1} for $m=1$,
and in Table \ref{tab:appendix_tab2} for $m=2$. 
A good agreement is found with our stability code, with a maximal  relative error that is less than $2\%$. 
Fig. \ref{fig:benchmark} (a,b) shows the real part of the radial velocity $\widehat{u}_r$, and of the azimuthal velocity $\widehat{u}_\theta$, respectively, Both velocity components compare well with Fig. 13(a) of \cite{yim2016stability}.

\begin{table} 
\centering
\setlength\tabcolsep{7pt}
\caption{Maximum growth rate and frequency comparisons between \cite{yim2016stability} and the present stability code for $m=1$, $\alpha=1.2$, ${Fr}=0.5$ and ${Re}=10^4$, and for different values of Rossby numbers. The \cite{yim2016stability} values are estimated from their Fig. (10).
}
\label{tab:appendix_tab1}
\scriptsize
\begin{tabular}{lr|rr}\toprule
\multirow{2}{*}{\small \text{Rossby number (Ro)}} &\multicolumn{2}{c}{\small $Ro\, \tilde{\omega}$} \\\cmidrule{2-3}
&{\small \text{Yim et. al (2016)}} & {\small \text{Present code}} \\\midrule
\small \text{$Ro=5$} &\multicolumn{1}{c}{$\approx$ \small $0.071-0.098 \ii$} 
&\multicolumn{1}{c}{\small $0.072-0.094 \ii$} 
\\
\small \text{$Ro=7.5$} &\multicolumn{1}{c}{$\approx$ \small $0.090-0.108 \ii$} 
&\multicolumn{1}{c}{\small $0.091-0.101 \ii$} 
\\
\small \text{$Ro=10$} &\multicolumn{1}{c}{$\approx$ \small $0.098-0.118 \ii$} 
&\multicolumn{1}{c}{\small $0.098-0.117 \ii$} 
\\
\bottomrule
\end{tabular}
\end{table}
\begin{table} 
\centering
\setlength\tabcolsep{7pt}
\caption{Maximum growth rate and frequency comparisons between \cite{yim2016stability} and the present stability code for $m=2$, $\alpha=1.2$, ${Fr}=0.5$ and ${Re}=10^4$, and for different values of Rossby numbers. The \cite{yim2016stability} values are estimated from their Fig. (15).
}
\label{tab:appendix_tab2}
\scriptsize
\begin{tabular}{lr|rr}\toprule
\multirow{2}{*}{\small \text{Rossby number (Ro)}} &\multicolumn{2}{c}{\small $Ro\, \tilde{\omega}$} \\ \cmidrule{2-3}
&{\small \text{Yim et. al (2016)}} & {\small \text{Present code}} \\ \midrule
\small {$Ro=5$} &\multicolumn{1}{c}{$\approx$ \small $0.017-0.233 \ii$} &\multicolumn{1}{c}{\small $0.016-0.236 \ii$} \\
\small {$Ro=7.5$} &\multicolumn{1}{c}{$\approx$ \small $0.011-0.233 \ii$} &\multicolumn{1}{c}{\small $0.012-0.235 \ii$} \\
\small {$Ro=10$} &\multicolumn{1}{c}{$\approx$ \small $0.008-0.233 \ii$}  &\multicolumn{1}{c}{\small $0.008-0.234 \ii$} 
\\
\bottomrule
\end{tabular}
\end{table}
\begin{figure}
    \centering
    \includegraphics[width=\textwidth]{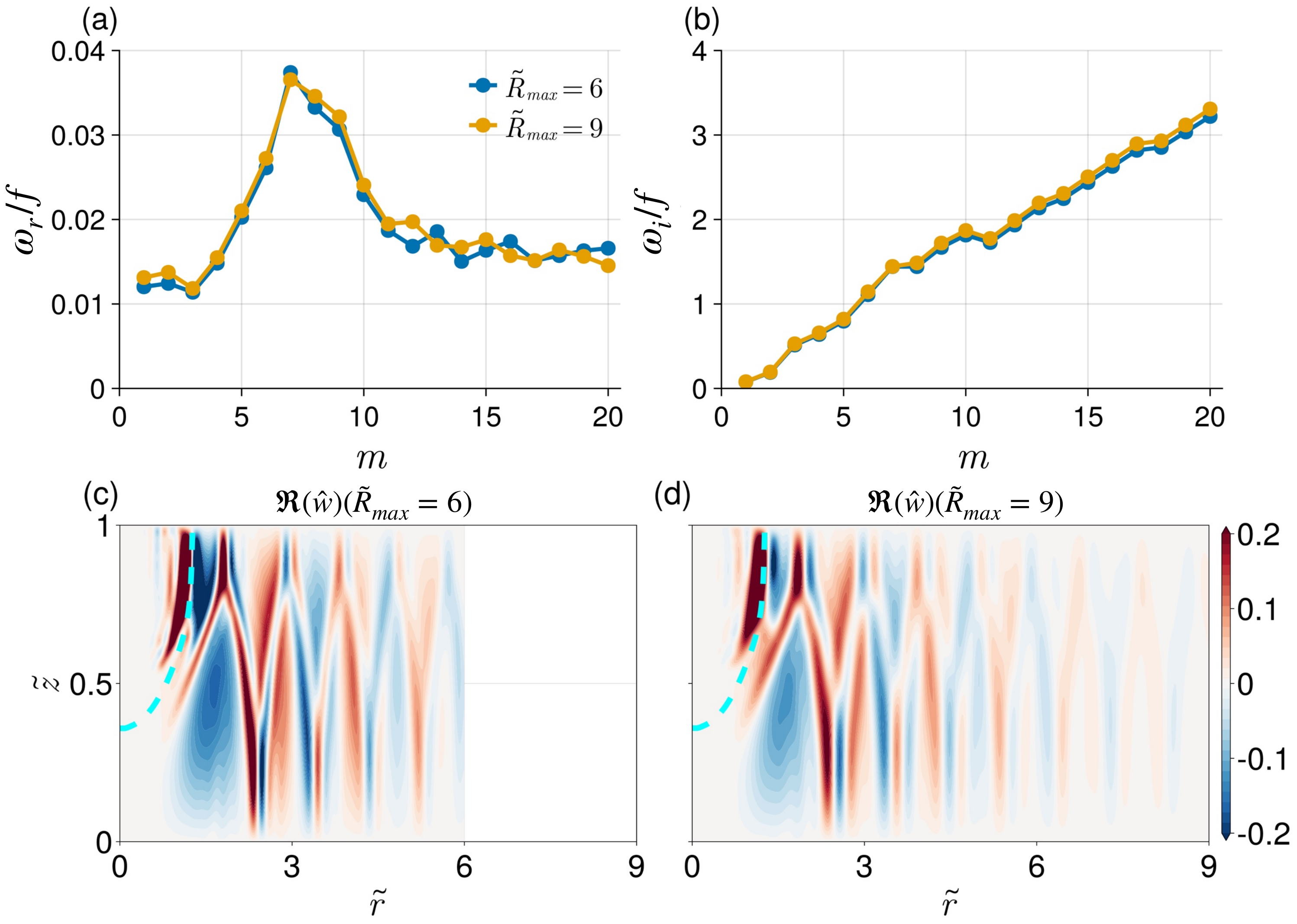}
    \caption{A comparison of (a) the nondimensional growth rate $\tilde{\omega}_r=\omega_r/f$ and (b) the nondimensional frequency $\tilde{\omega}_i=\omega_i/f$ for the  cases with $\tilde{R}_{max}=6,9$ (blue and yellow lines, respectively). Panel (c) and (d) show the real part of the vertical velocity eigenfunction $\mathfrak{R}(\hat{w})$ for the most unstable azimuthal wavenumber $m=7$, for $\tilde{R}_{max}=6$ and $\tilde{R}_{max}=9$, respectively. The cyan  line shows the critical radius $\tilde{r}_c(\tilde{z})$ given by 
    Eq. (\ref{critical_radii}). Note that the figure in panel (c) is plotted until $\tilde{r}=9$ for ease of comparison.
    }
    \label{fig:eigs_cmp}
\end{figure}

\appendix[B]
\appendixtitle{Stability analysis sensitivity to the radial domain size and number of grid points}
\label{sensitivity_check}
In this section we first test the sensitivity of the the linear stability analysis to the radial domain size $\tilde{R}_{max}$, by comparing two cases- $\tilde{R}_{max}=6$ and $\tilde{R}_{max}=9$. In both cases we use $N_z=80$ in the vertical and set $N_r/N_z=\tilde{R}_{max}$. The eigenvalues $\tilde{\omega}$ for different values of the azimuthal wavenumber $m$ are in good agreement in both cases (Figs. \ref{fig:eigs_cmp}(a,b)).  
Furthermore, the real part of the vertical velocity eigenfunction $\mathfrak{R}(\widehat{w})$, based on the most unstable mode $m=7$, exhibits similar structure in both cases (Figs. \ref{fig:eigs_cmp}(c,d)). This indicates that the results presented in the manuscript are converged for the maximal radial extend used ($\tilde{R}_{max}=9$).
\begin{figure}
    \centering
    \includegraphics[width=\textwidth]{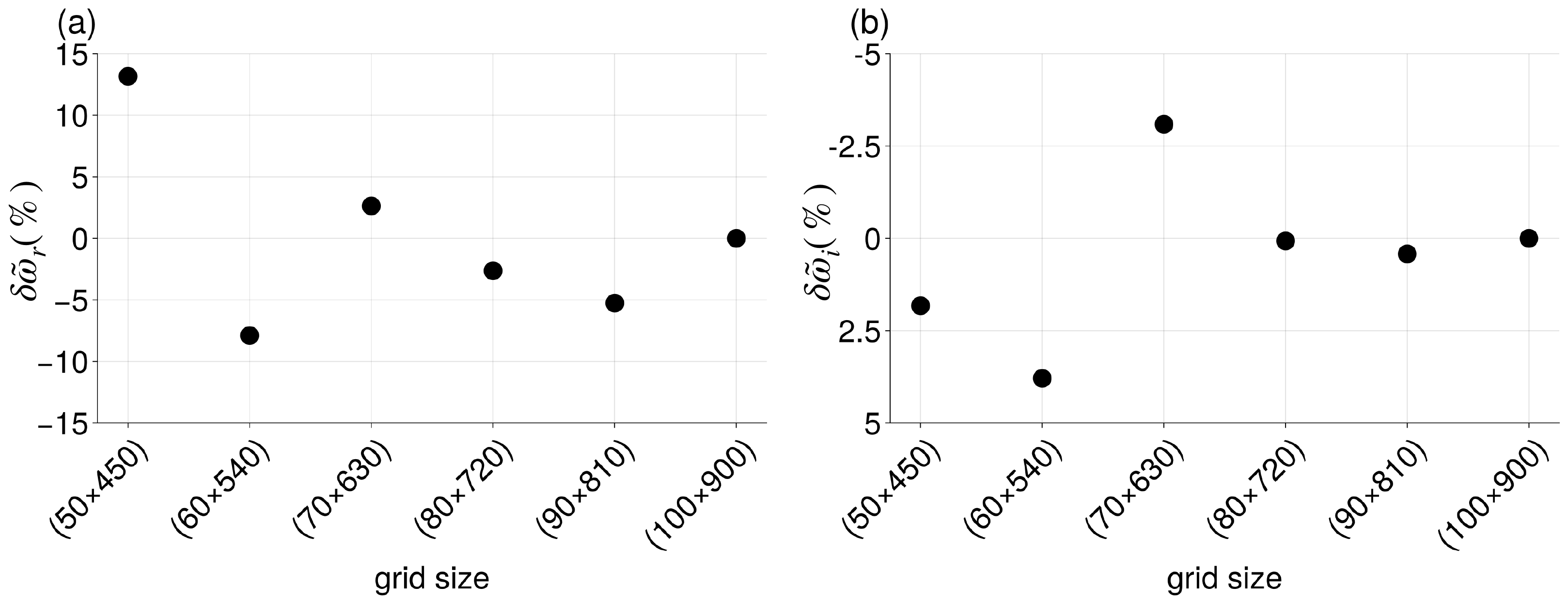}
    \caption{Grid convergence test for the linear stability analysis of the most unstable mode ($m=7$), and with $R_{max}=9$. Relative errors of (a) the growth rate $\tilde{\omega}_r$ and (b) frequency $\tilde{\omega}_i$ (Eqs. \ref{errors}(a,b)) are presented for different grid sizes $(N_z \times N_r)$. }
    \label{fig:grid_test}
\end{figure}

Second, we determine the grid resolution convergence for 
the most unstable mode, $m=7$, for the AAI case (Case 1 in Fig. \ref{fig:basic_state}). We vary the number of vertical grid points from ${N}_z = 50$ to $N_z=100$ while keeping the ratio $N_r/N_z=\tilde{R}_{max}$, where $N_r$ is the number of grid points in the $r$ direction. We consider the case with $\tilde{R}_{max}=9$,  which gives maximal matrix sizes ($\bm{\mathcal{A}}$ and $\bm{\mathcal{B}}$ in Eq. \ref{gen_eigvals}) of $450000^2$. We take the eigenvalue $\tilde{\omega}$ corresponding to $N_z=100$ as the ground truth and define the relative error of the growth rate $\tilde{\omega}_r$ and of the frequency $\tilde{\omega}_i$ to be 
\begin{subequations}
\label{errors}
    \begin{gather}
    \delta \tilde{\omega}_r(N_z) = \frac{\tilde{\omega}_r(N_z) - \tilde{\omega}_r(N_z=100)}{\tilde{\omega}_r(N_z=100)}, 
\,\,\,\,\,\,\,\
    \delta \tilde{\omega}_i(N_z) = \frac{\tilde{\omega}_i(N_z) - \tilde{\omega}_i(N_z=100)}{\tilde{\omega}_i(N_z=100)}.
    \tag{\theequation a-b}
    \end{gather}
\end{subequations}
For $N_z \geq 70$, we obtain a relative error of $ \leq 5\%$ for both $\tilde{\omega}_r$ and $\tilde{\omega}_i$ (Figs. \ref{fig:grid_test}(a,b)).
Results presented in this manuscript are therefore computed for $N_z=80$ and $\tilde{R}_{max}=9$.
\clearpage
\bibliographystyle{ametsocV6}
\bibliography{main}

\end{document}